\documentclass[12pt]{article}
\usepackage{amssymb}
\usepackage{a4wide}
\usepackage{times}
\usepackage{lineno}
\usepackage{caption}
\usepackage{subcaption}
\usepackage{amsmath}
\usepackage[export]{adjustbox}
\newsavebox\mybox
\usepackage[sort&compress,numbers]{natbib}
\usepackage[colorlinks=true,citecolor=blue]{hyperref}
\graphicspath{{Figures/}}
\title{A finite element analysis of rolling of bilayer films to cylindrical and conical tubes}
\author{Nihit Vyas and Ratna Kumar Annabattula\footnote{Corresponding author: ratna@iitm.ac.in}\\{\small{Department of Mechanical Engineering, Indian Institute of Technology Madras, Chennai-600036, India}}}
\begin{document}
\maketitle
	\section*{Abstract}
With recent developments in nanotechnology, self-assembled structures are providing convenient, cheaper and more precise ways of manufacturing various patterns and shapes with less complexity. Of these self-assembled structures, rolled-up nano-tubes play a vital role in various aspects. The notion is, the utilization of strain energy developed during epitaxial growth of a bilayer thin film over a substrate, mediated by a sacrificial layer. While the sacrificial layer is etched, the bilayer film is subjected to release its own in-built strain energy in the out-of-plane direction (3D structure) due to a bending stress induced by biaxial strain through the thickness, in the bilayer. This paper proposes a new method of fabricating conical self-rolled assembly by thickness and strain variations along the width of the bilayer and, cylindrical structure of variable radius due to thickness and strain variations along the length.\\
{\emph{Keywords}}: thin Film; bilayer; cone; finte element analysis\\
\section{Introduction}
\label{S:1}
Neoteric advances on self-assembly of various thin film structures\cite{schmidt,prinz,zhang2,schmidt2,zhang,chen,hamley,fedor} have prompted for proliferated research at nano-scale. Some of the prominent contributions in the field of self-assembled bilayer thin film systems (a double layered film) are nano-tubes\cite{schmidt,prinz,zhang2}, nano-scrolls\cite{zhang,chen,bell}, nano-rings\cite{schmidt2,zhang2}, etc.. It is a result of out-of-plane deformation of the bilayer due to a through thickness gradient of eigenstrain developed during the epitaxial growth of the film over the substrate. However, initially, the bilayer being strongly bonded to the substrate, the bilayer remains strained. The implication thus, is to overcome this bond energy, achieved by etching, and releasing the eigenstrains (due to lattice mismatch of film and substrate) in specific direction, in the form of rolling-up and/or wrinkling\cite{fedor} of the bilayer\cite{cendula}.
These structures contribute significantly to a wide variety of interest in fields of drug delivery\cite{hamley}, nano-tools\cite{solovev}, photonics\cite{kipp,li,quinones}, nano-jets and engines\cite{baraban,fennimore,mei,solovev2}, and thus, it makes a significant topic for exploration of different patterns caused by the interplay between geometry and strain, specially due to its precise assembly\cite{chen2}.
\\
Previously, simulation works have been explored along with experimental backings for radius estimates\cite{nikishkov2}, moving boundary analysis \cite{huang} to simulate progressive etching, bending and wrinkling mode preference\cite{cendula}, side of roll preference\cite{chun, alben}, twisting\cite{chen}, etc..
\\
In some of the recent works, cone shaped nanostructures have also been made\cite{fomin,gao,quinones} for the nano-jets and nano-motors. One of the methods is, by taking thickness gradient and/or strain gradient along the width, as observed in \cite{fomin}. Although not explicitly mentioned or worked upon previously, these gradients contribute to a lot of variations for the precise self-assembly of conical or coiled structures. 
\\
In this paper, a new method for fabricating conical bilayer structures is presented and, we further investigate the effect of geometric parameters (thickness, length and width) of the initial bilayer and strain variation (decoupled), on the final geometry, separately. Similarly, investigation of thickness and strain variations along the length of the bilayer, is also presented.
\section{Model setup}
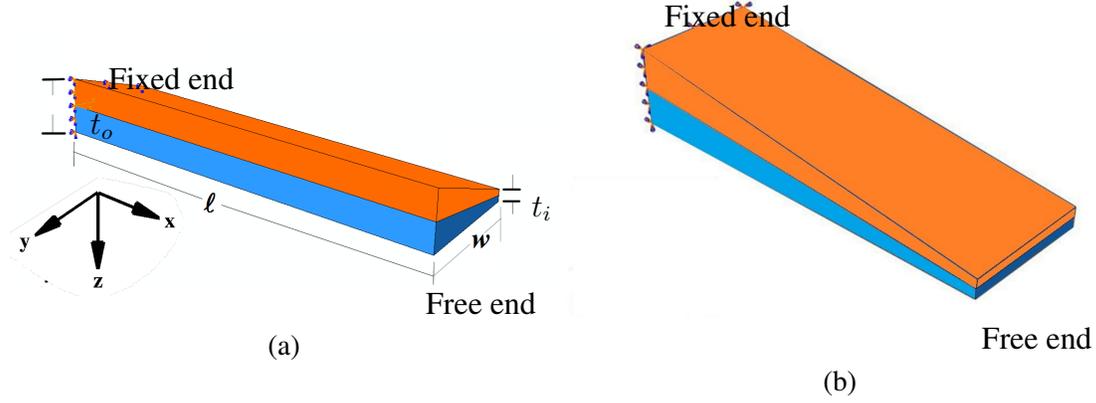
\begin{figure}[!htbp]
\centering
        \begin{subfigure}{0.45\textwidth}
                \includegraphics[width=\linewidth]{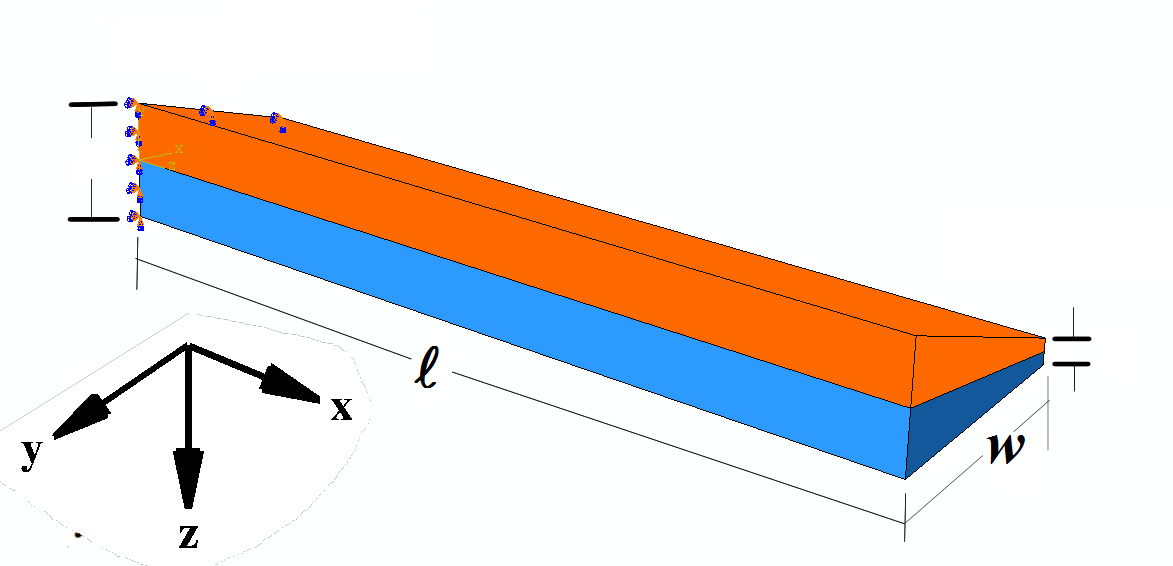}
                \put(-10,30){$t_i$}
                \put(-177,61){$t_o$}
                \put(-170,78){Fixed end}                            			\put(-50,-7){Free end}
                \caption{}
                \label{3a}
        \end{subfigure}
        \begin{subfigure}{0.45\textwidth}
                \includegraphics[width=\linewidth]{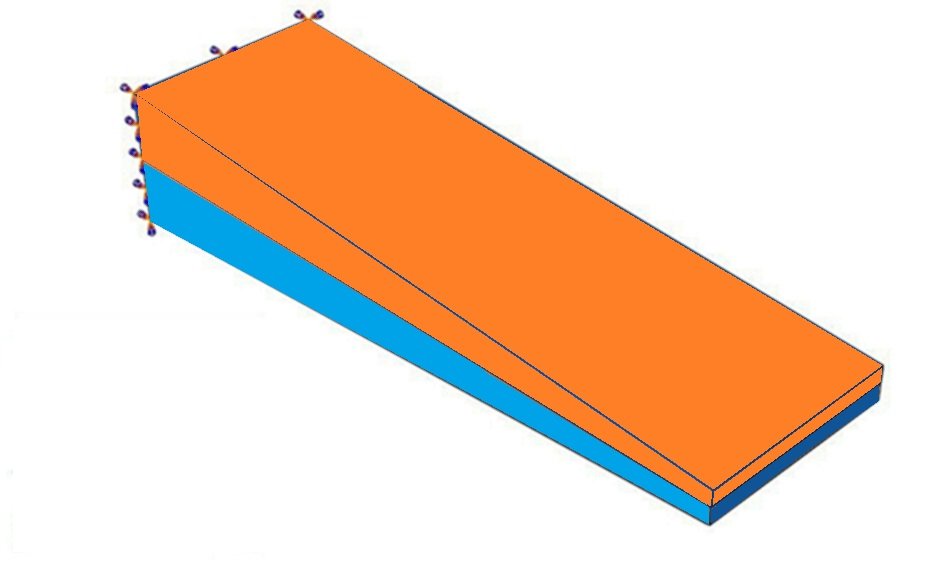}
                 \put(-170,115){Fixed end}                            				\put(-50,-7){Free end}
                \caption{}
                \label{3b}
        \end{subfigure}
	\caption{(a) Tapered thickness along the width and (b) Tapered thickness along the length in a bilayer system}\label{3}
\end{figure}

\begin{figure}[!htbp]
\centering
        \begin{subfigure}{0.49\textwidth}
                \includegraphics[width=\linewidth]{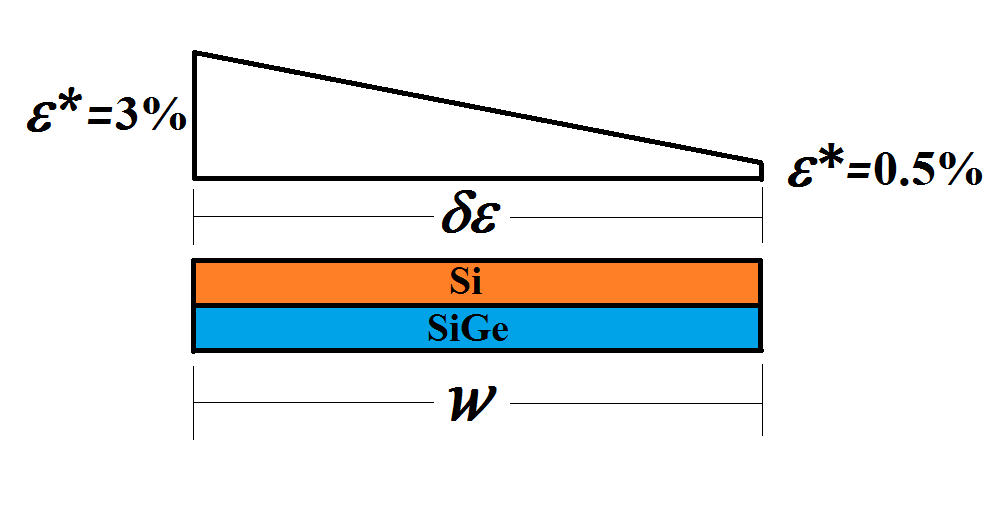}
                \caption{}
                \label{14a}
        \end{subfigure}
        \begin{subfigure}{0.49\textwidth}
                \includegraphics[width=\linewidth]{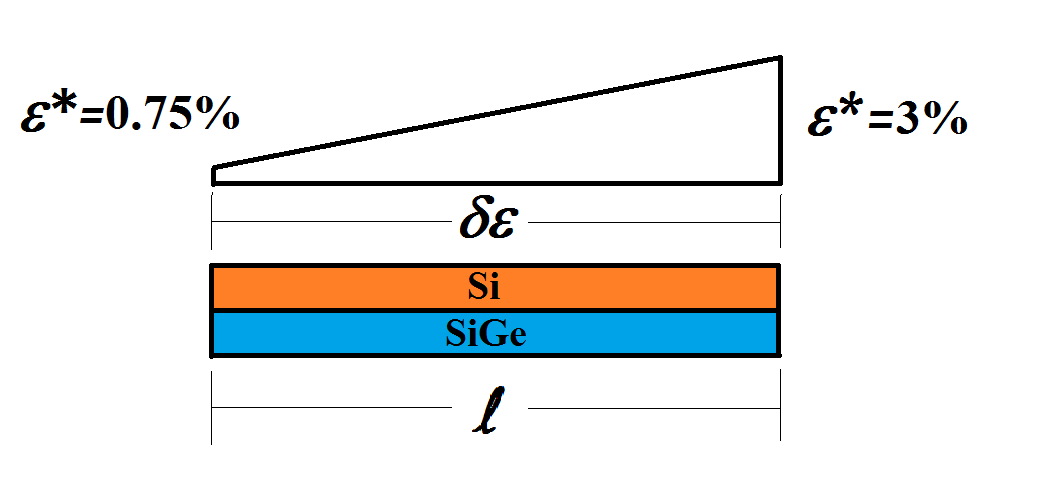}
                \caption{}
                \label{14b}
        \end{subfigure}
        \caption{a) Strain gradient along the width and b) Along the length in a bilayer system}\label{14}
\end{figure}

The formulation of a bilayer model is based on a multilayered plate system of Reissner-Mindlin plates\cite{reissner,mindlin} combined with F\"{o}ppl-von K\'{a}rm\'{a}n\cite{landau} equations for non-linear deformation, as the rolled-up assembly of the bilayer is subjected to small strains with large rotations. The roll-up process is simulated using Finite Element Method with the aid of commercial package ABAQUS\cite{abaqus}.
\noindent
The material properties (Young's modulus and Poisson's ratio) for the bilayer consisting of linear elastic isotropic Silicon (top layer) and Si$_x$Ge$_{1-x}$ (bottom layer) are taken as $E_1 = 170$ GPa, $\nu_1 = 0.22$\citep{sharpe} and $E_2 = 166.15$ GPa, $ \nu_2 = 0.2775$ ($x ~ 0.9$), respectively.\\
\noindent
In this section, thickness of a free-hanging bilayer is varied linearly, without any cohesive layer or substrate as shown in Fig.~\ref{3}. Also, strain in the Si$_{x}$Ge$_{1-x}$ layer is varied separately for a uniformly thick (both the layers of same non-varying thickness) free-hanging bilayer as shown in Fig.~\ref{14}.\\
The model consists of a bilayer, fixed at left end of the longer edge along the width and free at the other, with length $l=$ 20 $\mu$m and a linearly tapered, symmetric thickness cross-section, for thickness gradient along the width ($w$) (Fig.~\ref{3a}) and along the length ($l$) (Fig.~\ref{3b}). Eigenstrain is fixed at $2\%$ for all thickness variation analysis. For the thickness variation along the width, we first keep the taper ratio ($t_o/t_i$) constant, varying only the width, to simulate the effects of taper angle (the angle formed by the thickness gradient) by varying aspect ratio ($w/l$). Here, $t_i$ and $t_o$ are the total thickness of bilayer at small thickness and large thickness edge, respectively.\\
 The effect of the variation of taper angle is further studied, by keeping $t_i$ and $w/l$ constant and varying $t_o$. Taper angle, thus, is varied in terms of  taper ratio.\\
Finally, the effect of average thickness ($(t_o +t_i)/2$) for same taper angle/taper ratio was studied by keeping initial taper ratio as 5 and then adding 10 to 40 nanometers to the whole bilayer, terming this as average thickness increment, uniformly by maintaining proportionality in thickness of the layers as (1:1) in the bilayer. Width
is also varied for each average thickness case, for the complete study.

\noindent
Similarly, for strain gradient ($0.75\%$ to $3\%$) along the length and along the width ($0.5\%$ to $3\%$), bilayer with uniform cross section (with 30 nm thickness of a layer) is used with equal thickness for both layers. Minimum strain along the length was chosen as $0.75\%$ so as to observe higher deformation/rotation (also, to overcome the effects of fixed boundary). Likewise, minimum strain of $0.5\%$ is chosen for gradient along the width so as to avoid strain free condition and provide some bending moment at the respective edge.
\\

Rolled-up cylindrical nano-tubes have certain radius ($R$), which depends on the ratio of thickness ($t_1/t_2$) of the individual layers in a bilayer, and on strain difference in upper (subscript 1) and lower layer (subscript 2), $\delta \epsilon=\epsilon_1-\epsilon_2$ ($\epsilon$ represents strain), and material modulus ratio $\chi= Y_1/Y_2$, according to \cite{nikishkov}
\begin{equation}
R = 	\frac{Y^2_1 t^4_1+ Y^2_2 t^4_2 + 2 Y_1 Y_2 t_1 t_2 (2t^2_1+2t^2_2+3t_1t_2)}{6 Y_1 Y_2 t_1 t_2 (t_1+t_2)(\eta_1\epsilon_1-\eta_2\epsilon_2)}
\end{equation}
where, $Y_1=E_1/(1-\nu^2_1), Y_2=E_2/(1-\nu^2_1)$ and $\eta_{i}=1$, for plane strain condition.
\begin{figure}[!htbp]
\centering
        \includegraphics[width=0.5\linewidth]{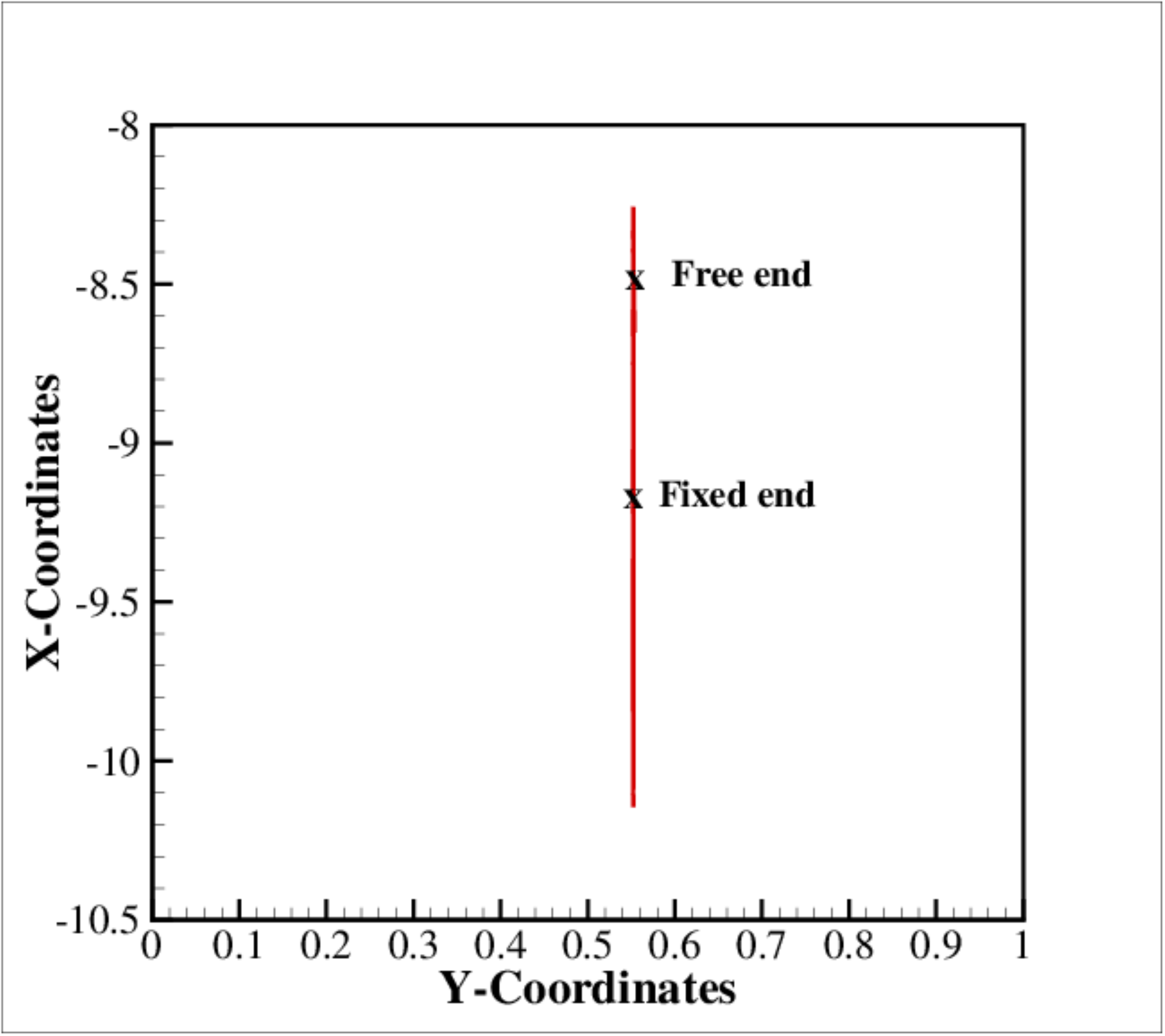}
                \caption{Projection of trace of a longitudinal edge of a sample rolled up uniform bilayer of certain uniform strain in lower layer (S$i_{x}$Ge$_{1-x}$ layer) and no strain in upper layer (Si layer) on to $xy$-plane.}
                \label{new}
\end{figure}
The projection of trace of a longitudinal edge of rolled-up bilayer, initially in $xy$-plane before rolling-up, on to the  $xy$ - plane is observed to be a straight line, as in Fig.~\ref{new}. However, this deviates for the thickness variation of bilayer along the width and strain gradient along the width, as discussed below.

\subsection{Effects of thickness variations along the width of a free standing bilayer film}
\begin{figure}[!htbp]
\centering
        \begin{subfigure}{0.49\textwidth}
                \includegraphics[width=\linewidth]{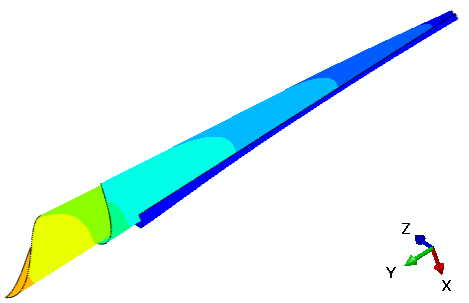}
                \caption{}
                \label{4a}
        \end{subfigure}
        \begin{subfigure}{0.49\textwidth}
                \includegraphics[width=\linewidth]{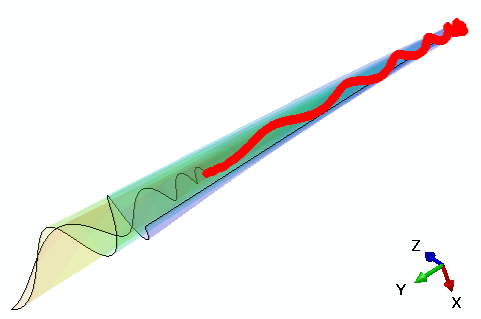}
                \caption{}
                \label{4b}
        \end{subfigure}
        \begin{subfigure}{0.49\textwidth}
                \includegraphics[width=\linewidth]{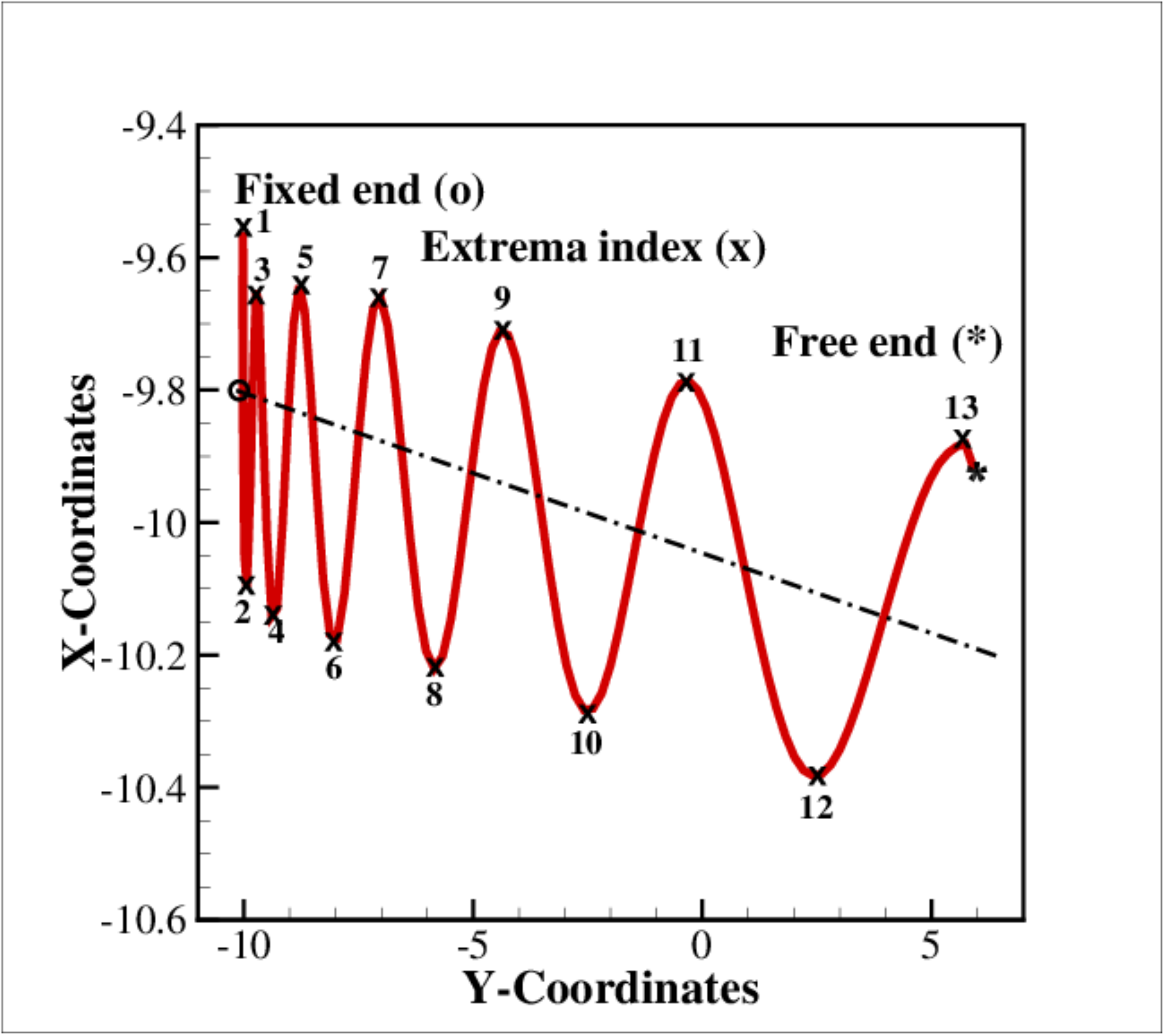}
                \caption{}
                \label{4c}
        \end{subfigure}
	\hfill
        \begin{subfigure}{0.49\textwidth}
                \includegraphics[width=\linewidth]{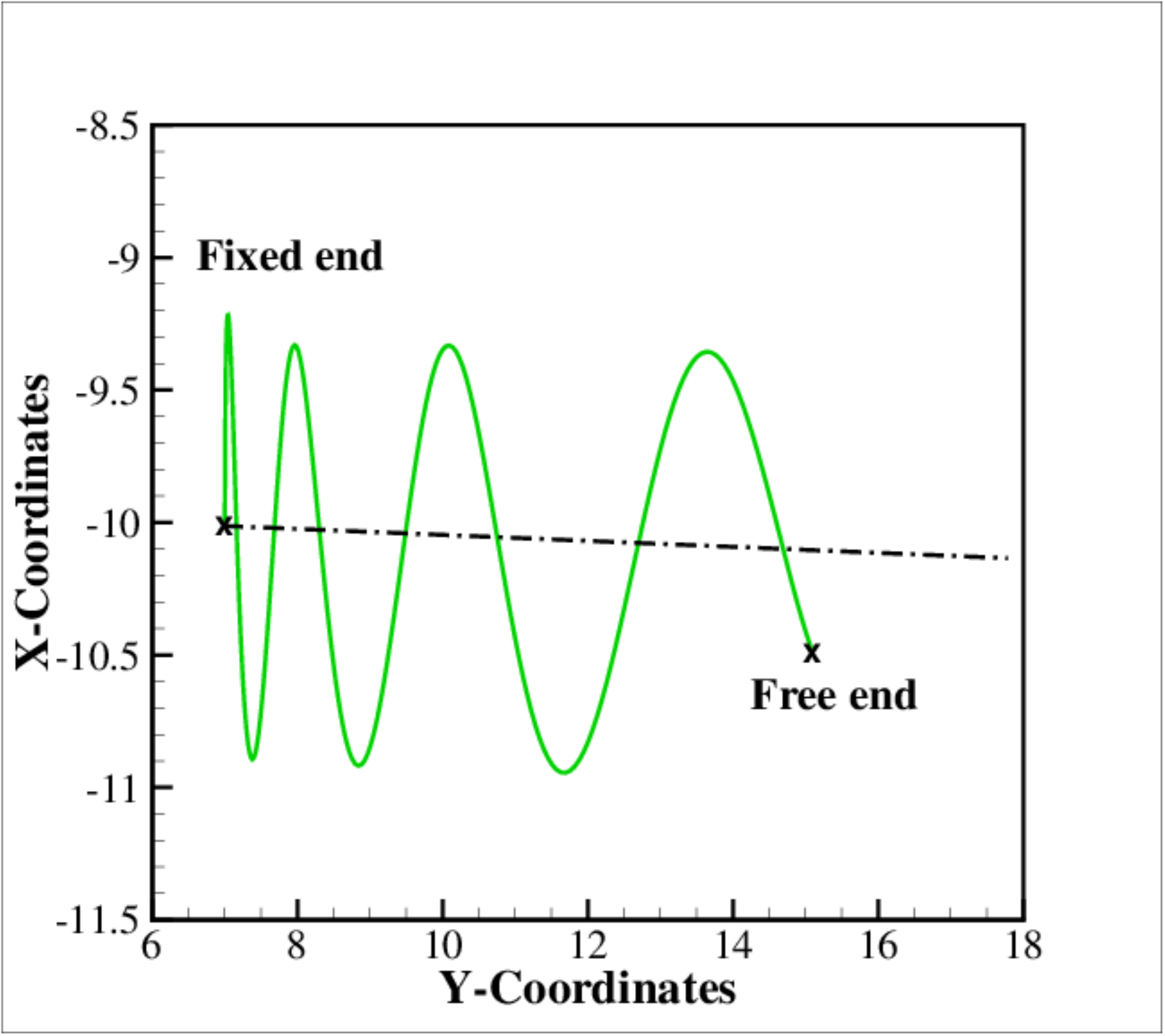}
                \caption{}
                \label{4d}
        \end{subfigure}
	\caption{a) Conical structure formed by thickness gradient along the width, b) Tracing the deformed small thickness edge, c) Projection of small thickness edge on $xy$-plane and d) Projection of large thickness edge on $xy$-plane.}\label{4}
\end{figure}
In this section, we study the effect of geometry, i.e., taper angle (by taper ratio ($t_o / t_i$)), aspect ratio ($w/l$) variation (by varying width) and average thickness ($(t_o + t_i)/2$) variation (same taper ratio, adding overall thickness), on the cone angle, pitch and radius of the rolled-up tube. Due to the asymmetry introduced through tapered cross section, the film also bends in $y$-direction, as shown in Fig.~\ref{4}. This leads to formation of a helical roll with increasing pitch, as a direct consequence of varying bending stiffness along the width. Conical shapes due to thickness gradient along the width, have been previously observed experimentally\cite{fomin}. However, there was no systematic study on such conical system in literature to the best of our knowledge.
\par\noindent
We study the change in pitch, for different aspect ratio ($w/l$), taper ratio ($t_o / t_i$) and average thickness, between same extrema number, starting from fixed end to free end as depicted in Fig.~\ref{4c}. All the variations were studied by tracing and projecting the deformed state of small thickness edge and large thickness edge on the $xy$-plane as shown in Fig.~\ref{4c} and ~\ref{4d}, respectively.\\
\subsubsection{Effects of width variation}
\begin{figure}[!htbp]
\centering
        \begin{subfigure}{0.49\textwidth}
                \includegraphics[width=\linewidth]{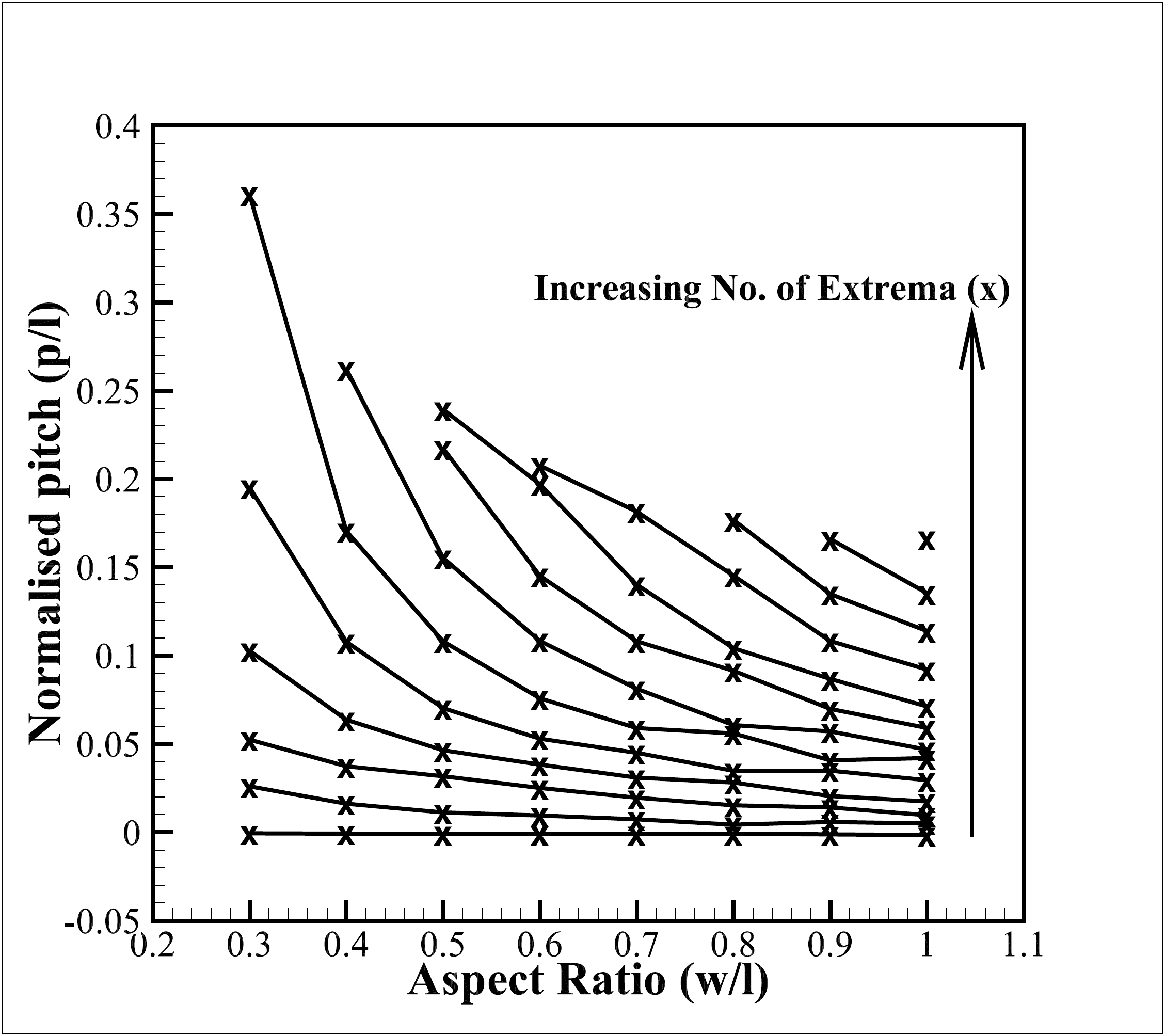}
                \caption{}
                \label{6a}
        \end{subfigure}
        \hfill
        \begin{subfigure}{0.49\textwidth}
                \includegraphics[width=\linewidth]{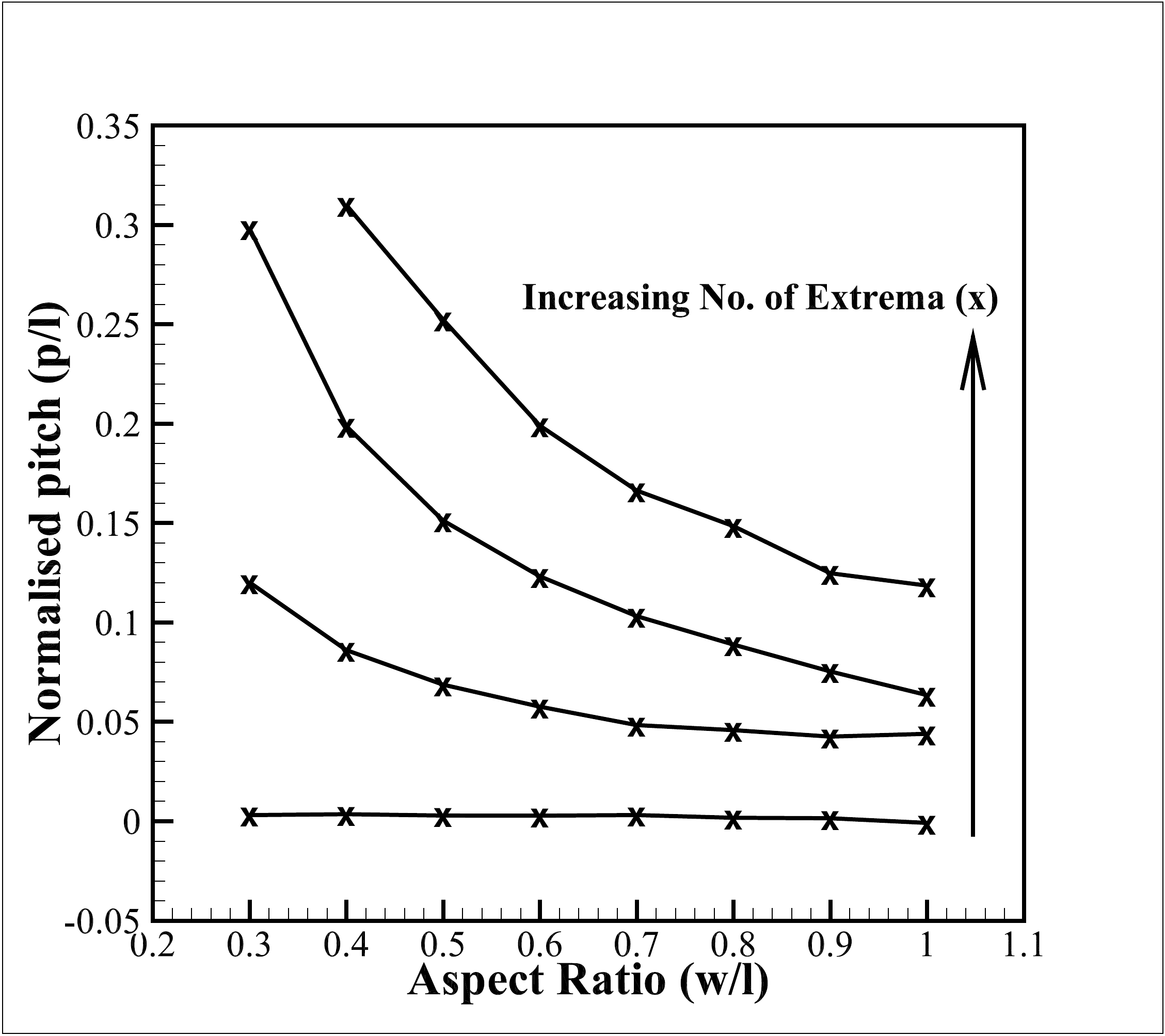}
                \caption{}
                \label{6b}
        \end{subfigure}
        \caption{ Aspect ratio effect:  a) Pitch variation plot for small thickness edge and b) For large thickness edge}\label{6}
\end{figure}
\begin{figure}[!htbp]
\centering
        \begin{subfigure}{0.49\textwidth}
                \includegraphics[width=\linewidth]{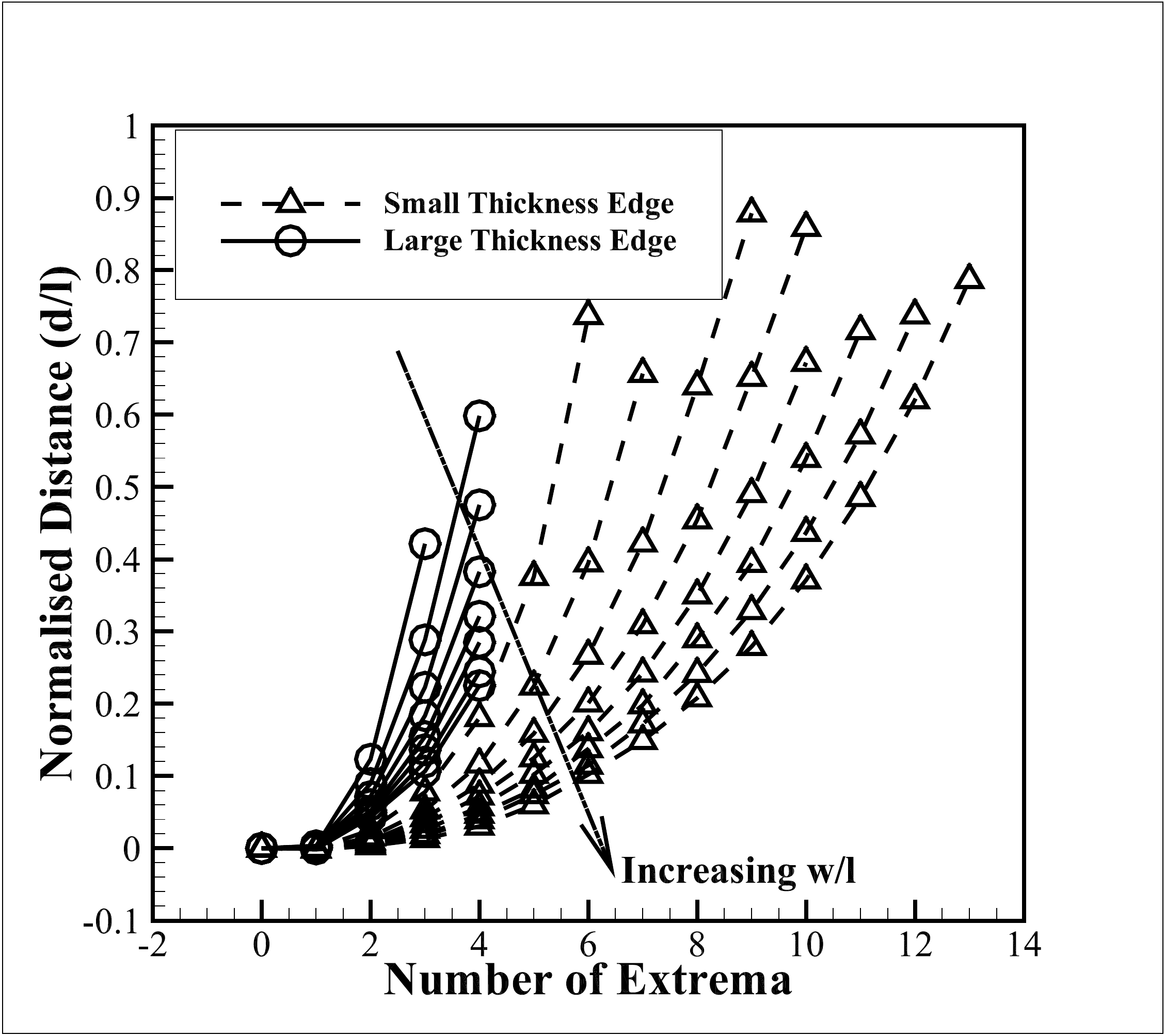}
                \caption{}
                \label{7a}
        \end{subfigure}
        \hfill
        \begin{subfigure}{0.49\textwidth}
                \includegraphics[width=\linewidth]{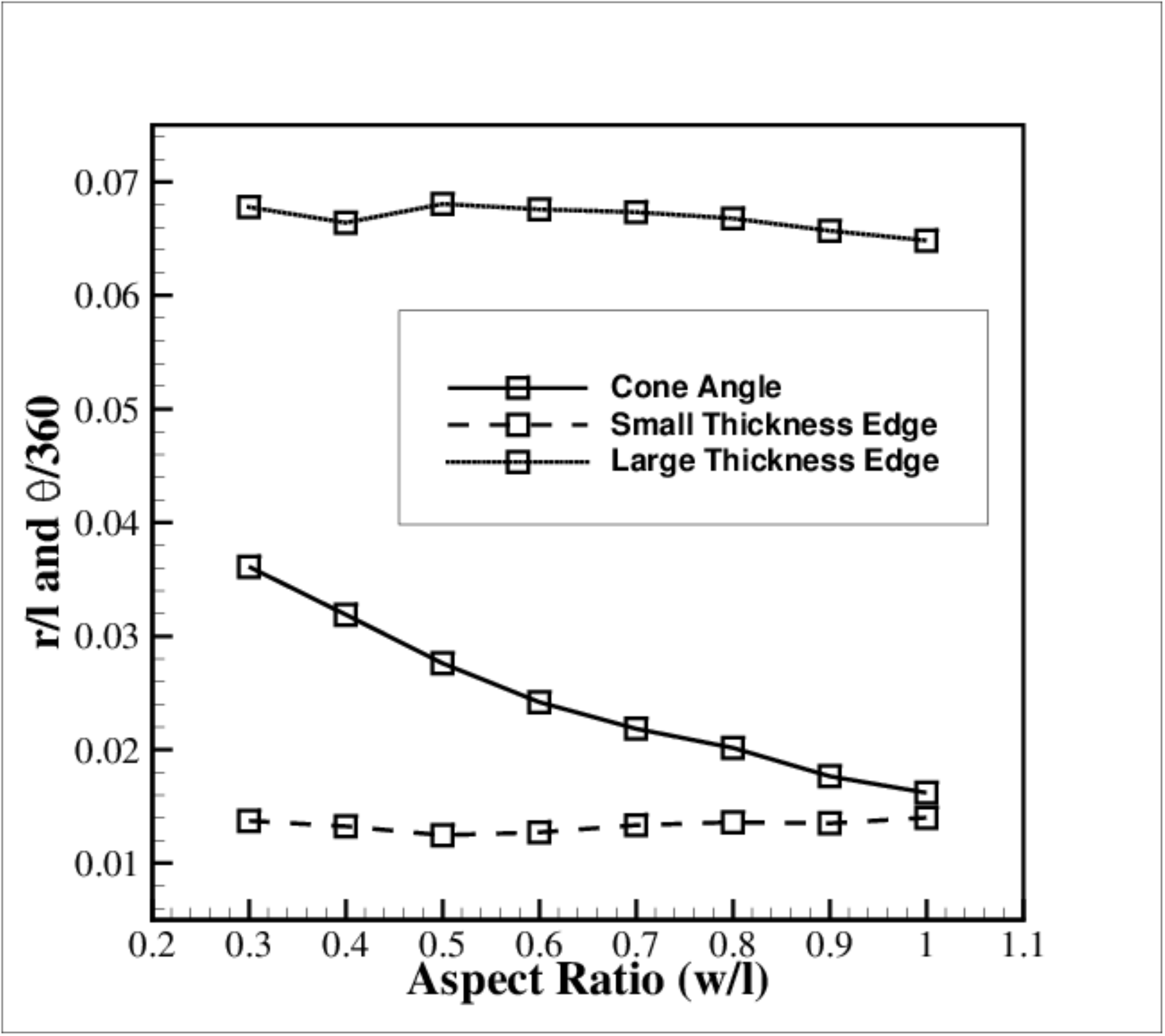}
                \caption{}
                \label{7b}
        \end{subfigure}
        \caption{ Aspect ratio effect: a) Location of an extrema along median wave-line path $d$ for different aspect ratios ($w/l$) from fixed end to free end of an edge's projection plot and b) Normalised radius ($r/l$) (non-solid lines) and normalised cone angle}\label{7}
\end{figure}

It can be observed from Figs.~\ref{6} and ~\ref{7a}, as the aspect ratio $w/l$ increases, number of turns increase for both small thickness edge and large thickness edge, resulting in formation of a compact conical roll. Fig.~\ref{6} suggests, more the aspect ratio ($w/l$), lesser the pitch ($p$) becomes, between two extrema number (read as a curve in Fig.~\ref{6}), i.e., for a given turn number, for both small thickness edge and large thickness edge. It may further be noted that the increase in the number of turns with increase in the aspect ratio ($w/l$) is quantitatively larger for the small thickness edge than the large thickness edge. This is due to lower bending stiffness of the small thickness edge compared to the large thickness edge. While the radius of any given section along the length of the bilayer remains same, it linearly increases from small thickness edge (smallest radius) to large thickness edge (largest radius), forming a constant cone angle, as shown in Fig.~\ref{7b}, which decreases as the aspect ratio ($w/l$) increases. Thus, supporting the previous statement of compactness, as indicated by Fig.~\ref{6}. Fig.~\ref{7b} also indicates that the radius of any longitudinal section of the bilayer is independent of the aspect ratio ($w/l$). Note that the thickness of the film changes at a longitudinal section with width variation for a given $t_o$ and $t_i$. Hence, it may be concluded that for a given thickness the radius remains independent of the aspect
ratio ($w/l$). However, this radius can't be obtained from equation (1) due to the effects
of thickness variation along width.\\
\subsubsection{Effects of taper ratio ($t_o / t_i$)}
\begin{figure}[!htbp]
\centering
        \begin{subfigure}{0.49\textwidth}
                \includegraphics[width=\linewidth]{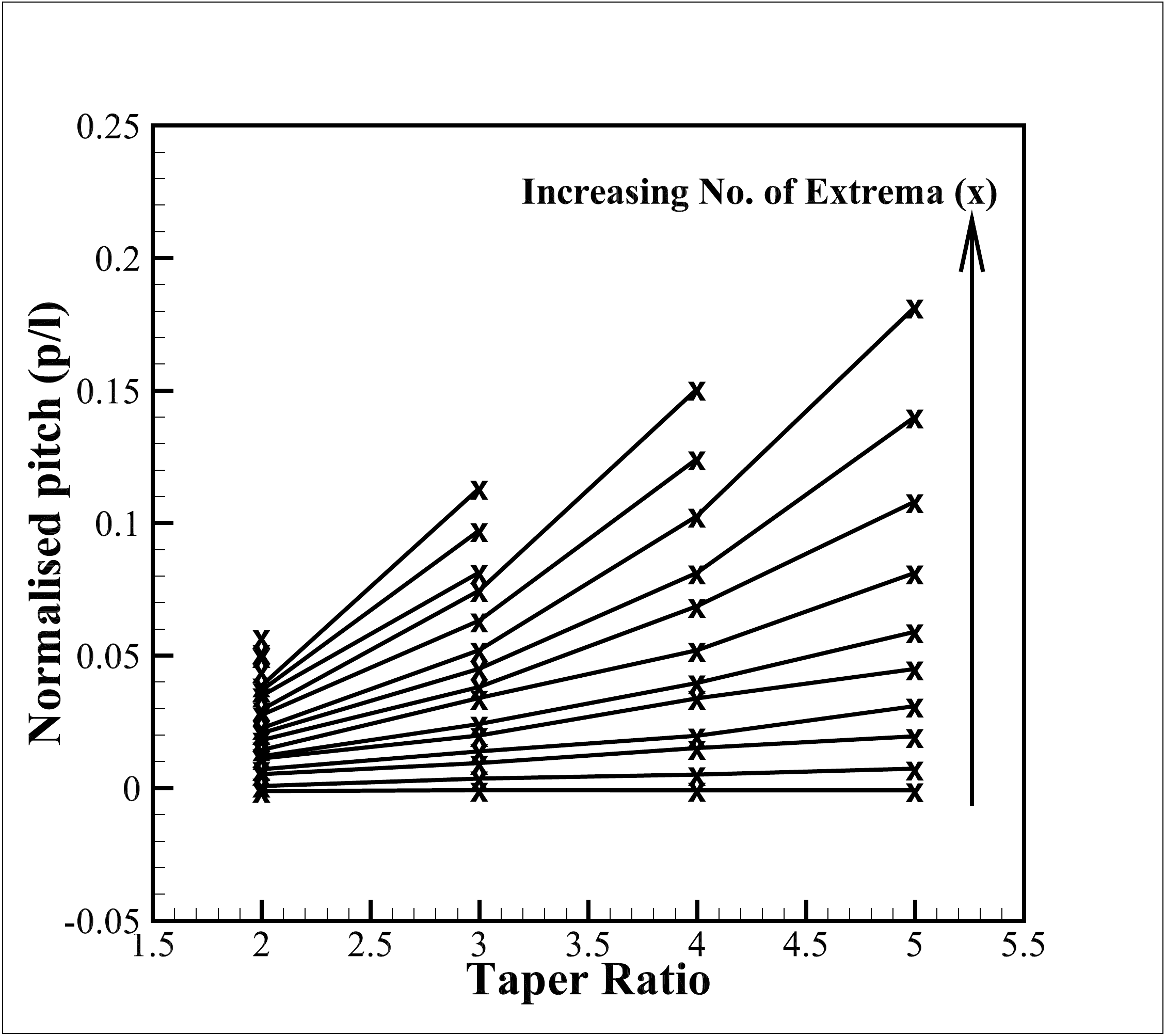}
                \caption{}
                \label{8a}
        \end{subfigure}%
        \hfill
        \begin{subfigure}{0.49\textwidth}
                \includegraphics[width=\linewidth]{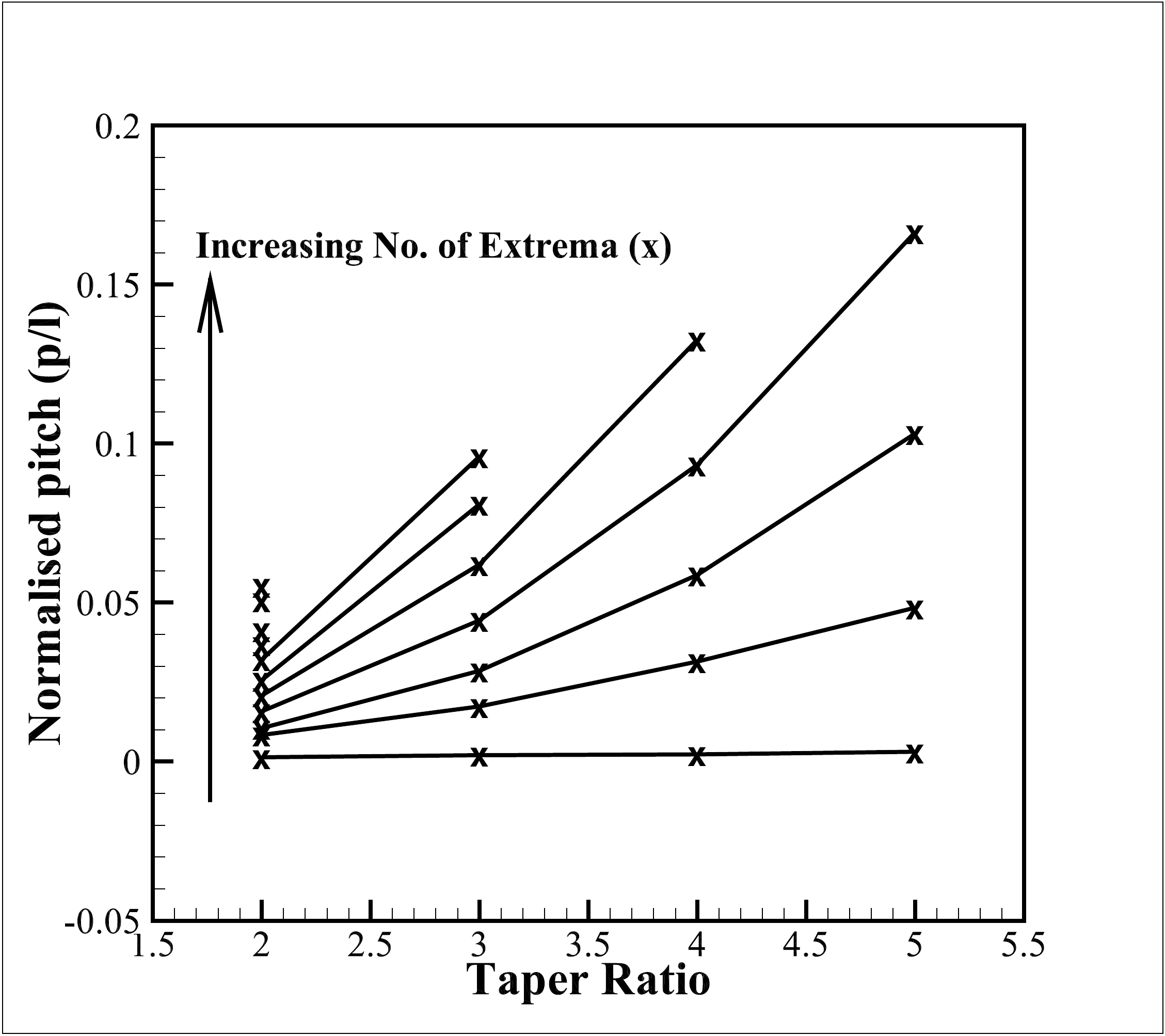}
                \caption{}
                \label{8b}
        \end{subfigure}%
        \hfill
        \begin{subfigure}{0.49\textwidth}
                \includegraphics[width=\linewidth]{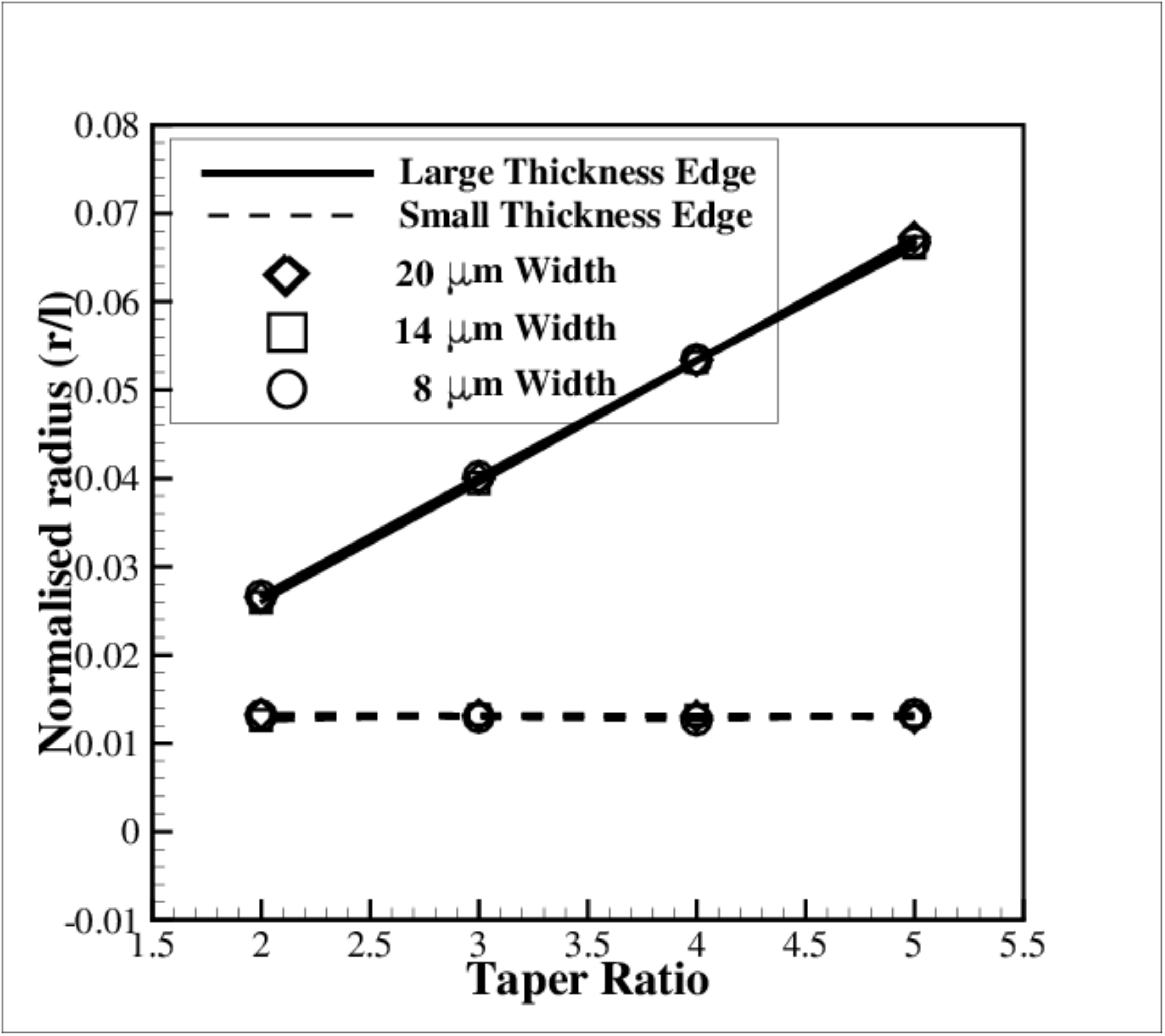}
                \caption{}
                \label{8c}
        \end{subfigure}%
        \hfill
               \begin{subfigure}{0.49\textwidth}
                \includegraphics[width=\linewidth]{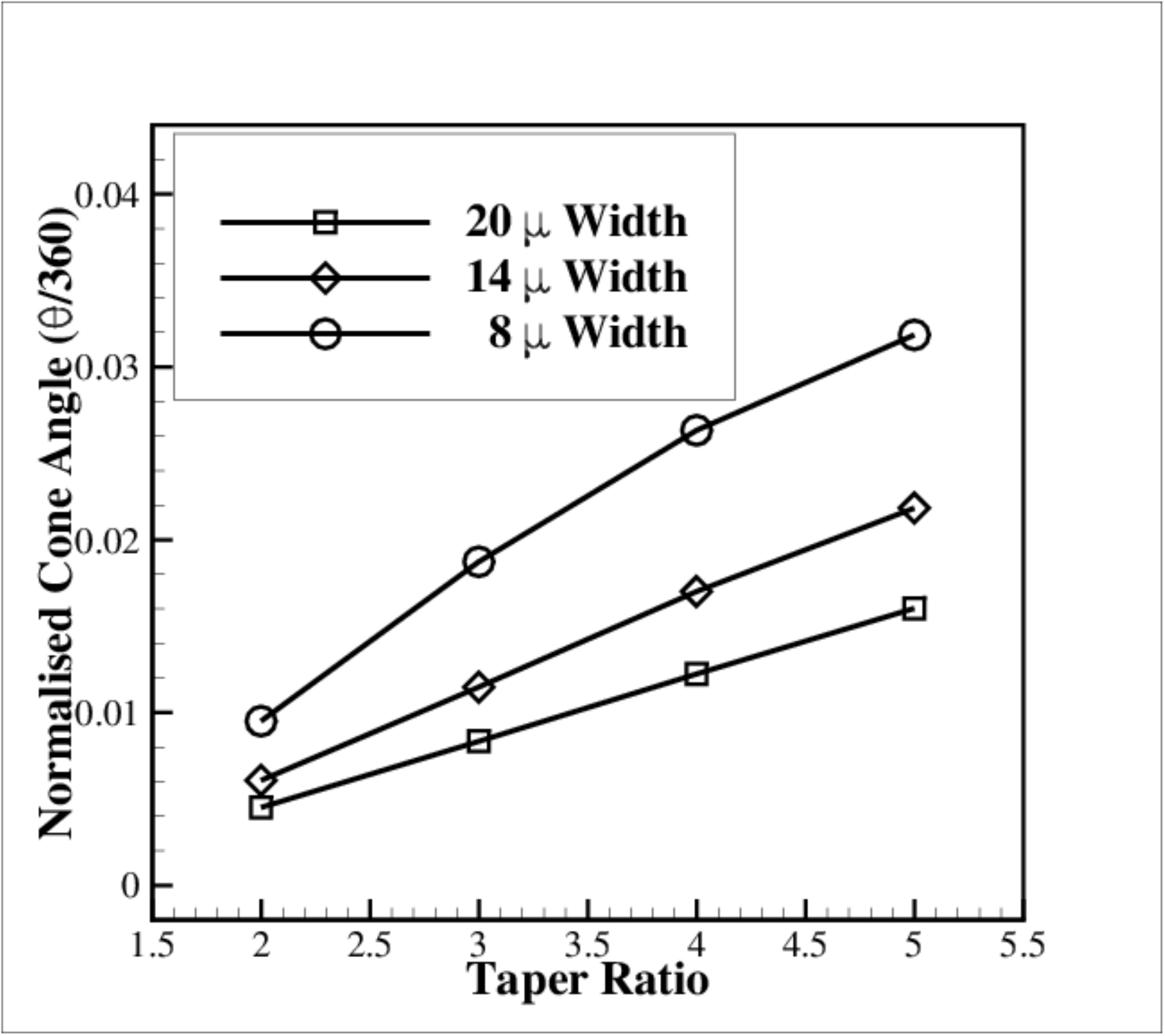}
                \caption{}
                \label{8d}
        \end{subfigure}%
        \caption{Taper ratio effect: a) Pitch variation plot for small thickness edge, b) For large thickness edge, c) Normalised radius ($r/l$) and d) Normalised cone angle}\label{8}
\end{figure}

Fig.~\ref{8a} and~\ref{8b} show the effect of varying taper ratio ($t_{o}$/$t_{i}$) for a given width of the bilayer, on the geometry of conical roll. The value of $t_i$ is taken as 10 nm while the value
of $t_o$ is varied. As the taper ratio increases (See Fig.~\ref{8a} and ~\ref{8b}), the number of turns decreases, and hence, the pitch between two particular extrema increases. As observed and stated previously, the radius
for the same thickness remains the same and increases with increasing thickness at the large thickness edge (as thickness of outer edge is varied) as observed in Fig.~\ref{8c}. The overlapping of curves for different widths for small thickness edge reiterates the prediction. The cone angle increases with increasing the taper ratio ($t_{o}$/$t_{i}$) as shown in Fig.~\ref{8d}, and decreases with increase in aspect ratio, for a fixed taper ratio, as observed previously in Fig.~\ref{7b}.
\subsubsection{Effects of average thickness variation}
\begin{figure}[!htbp]
\centering
        \begin{subfigure}[t]{0.49\textwidth}
                \includegraphics[width=\linewidth]{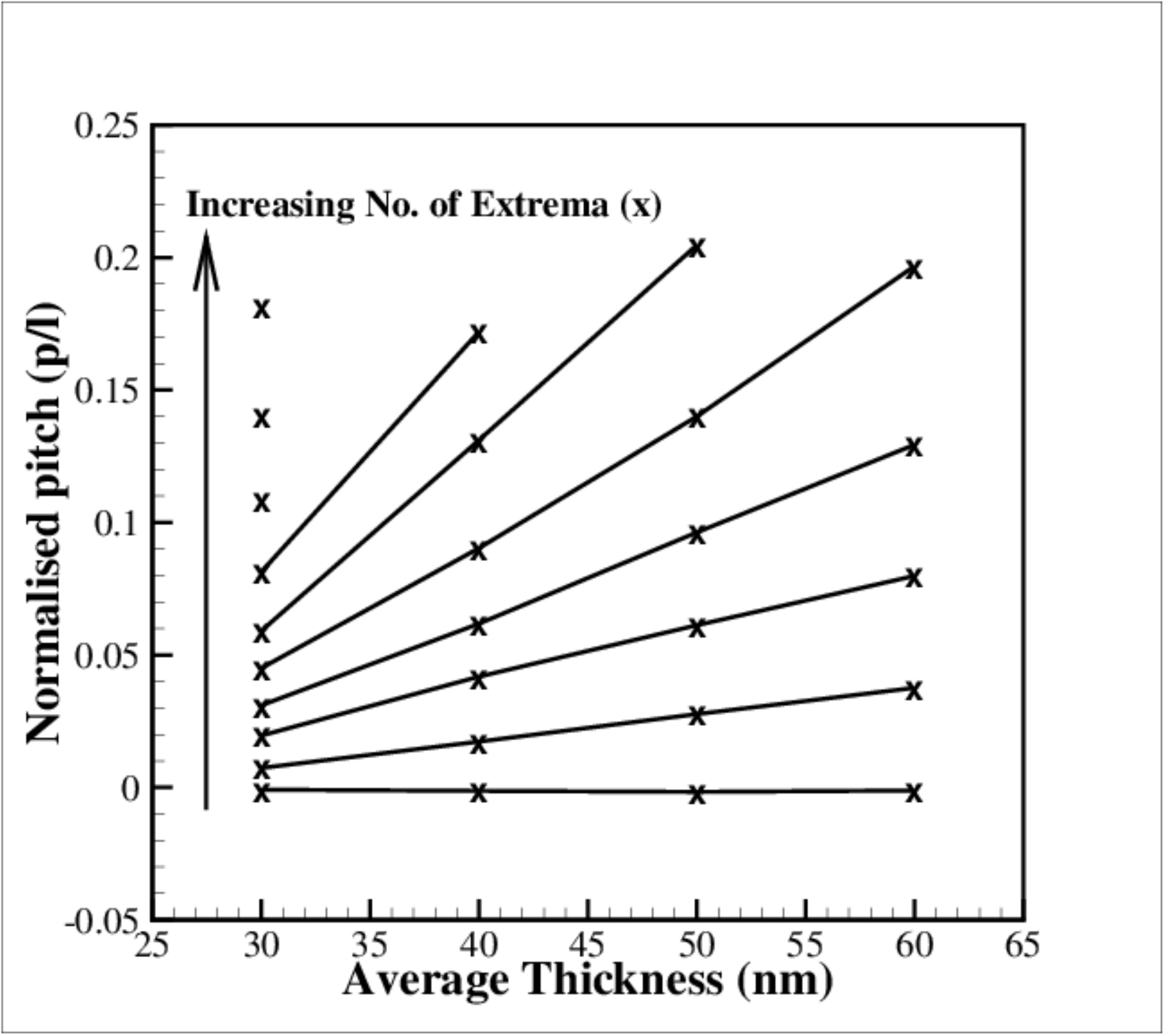}
                \caption{}
                \label{9a}
        \end{subfigure}%
        \hfill
        \begin{subfigure}[t]{0.49\textwidth}
                \includegraphics[width=\linewidth]{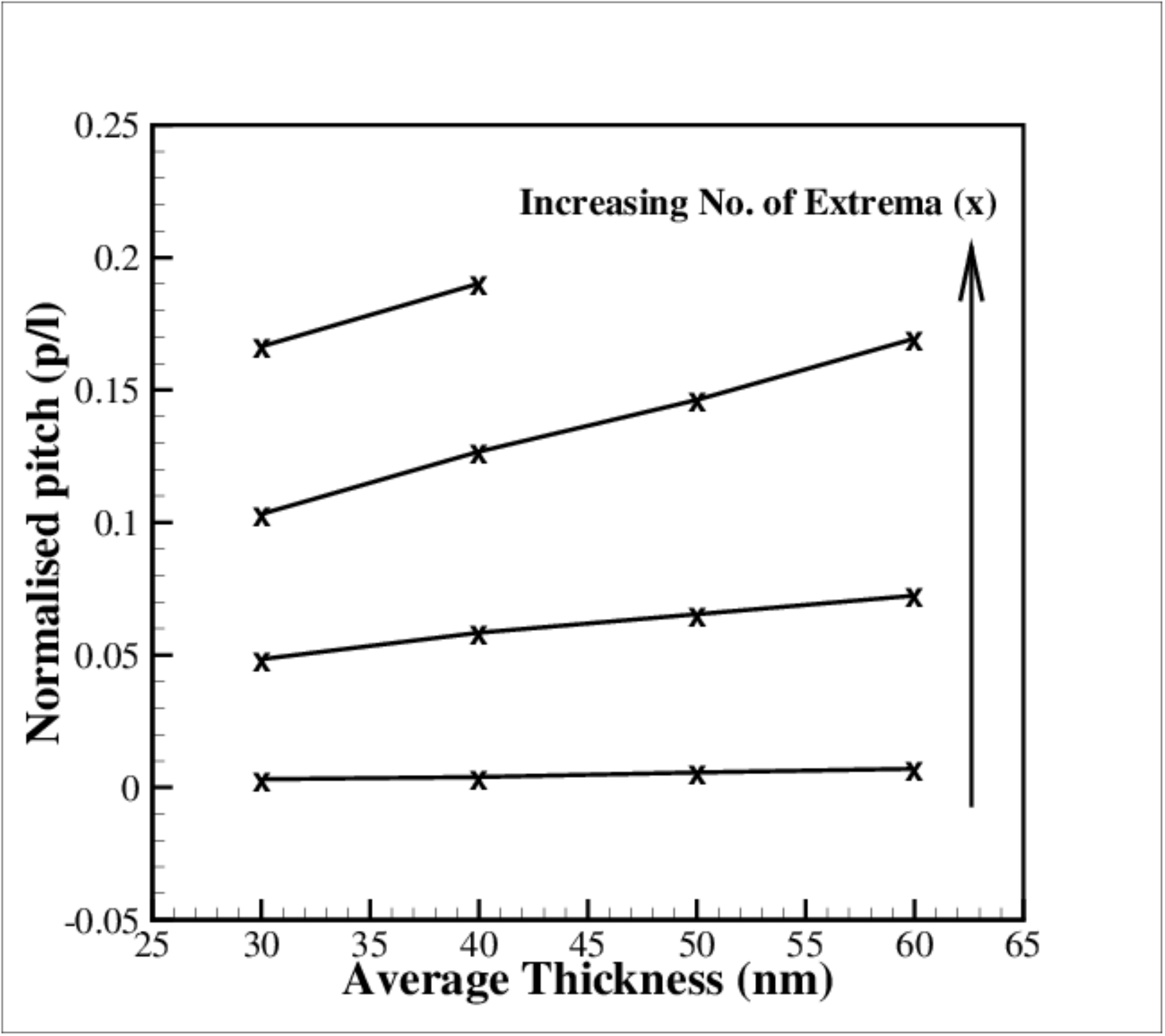}
                \caption{}
                \label{9b}
        \end{subfigure}%
        \hfill
        \begin{subfigure}[t]{0.49\textwidth}
                \includegraphics[width=\linewidth]{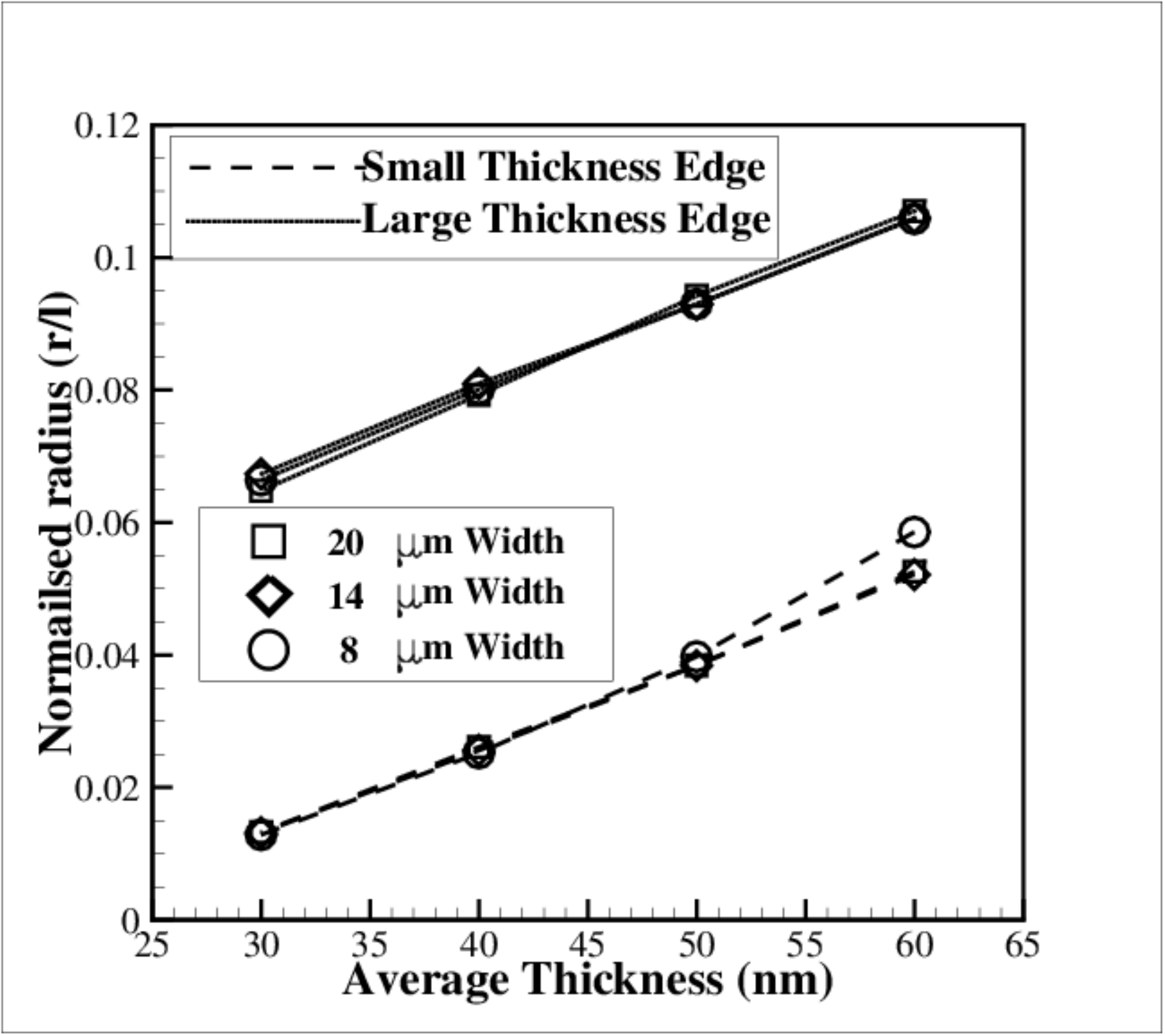}
                \caption{}
                \label{9c}
        \end{subfigure}%
        \hfill
                \begin{subfigure}[t]{0.49\textwidth}
                \includegraphics[width=\linewidth]{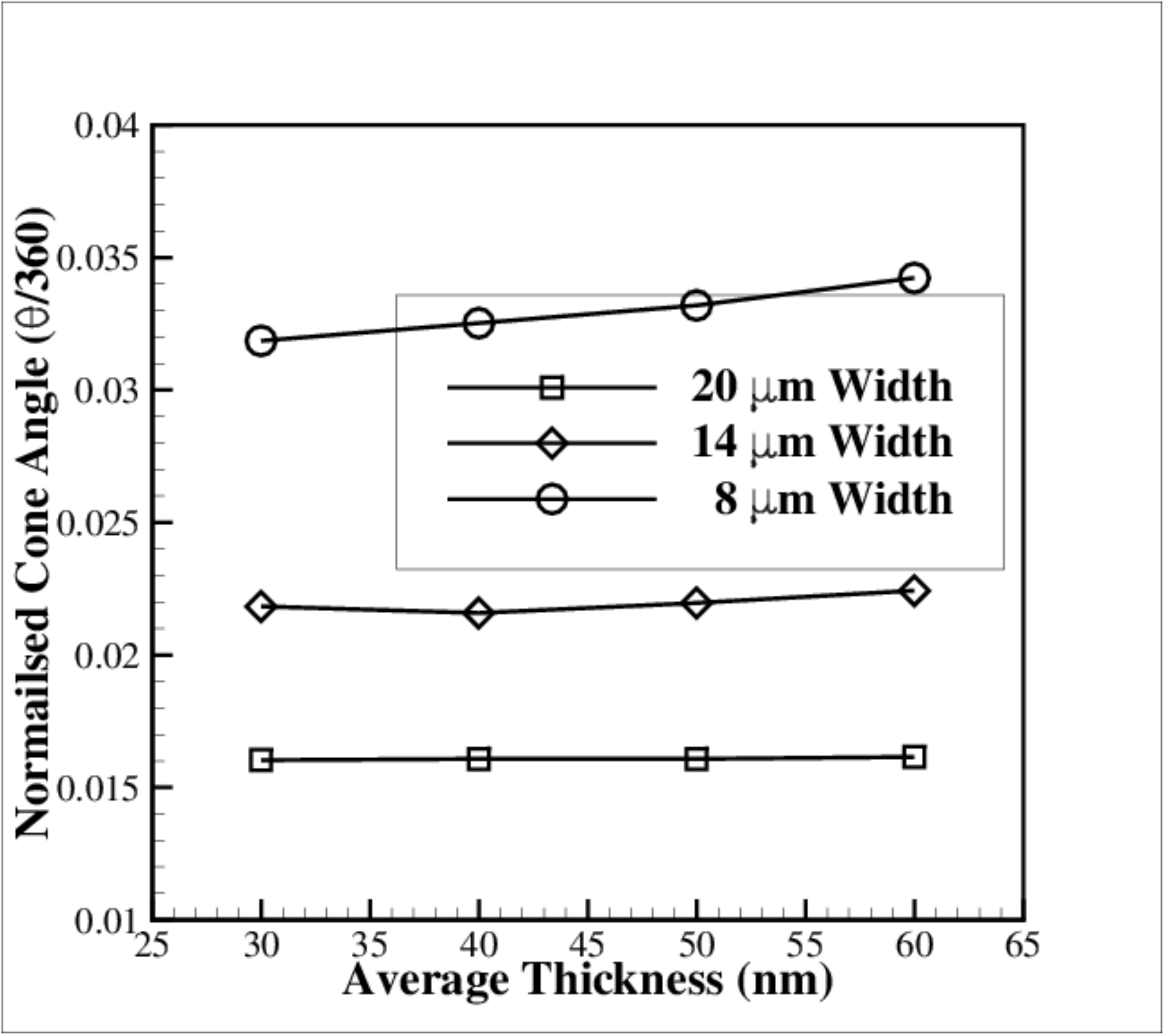}
                \caption{}
                \label{9d}
        \end{subfigure}%
        \caption{Average thickness effect (taper angle fixed): a) Pitch variation plot for small thickness edge, b) For large thickness edge, c) Normalised radius ($r/l$) (non-solid lines) and d) Normalised cone angle}
        \label{9}
\end{figure}

In this section, we study the effect of increasing the film thickness at both the large thickness edge and small thickness by the same amount thus keeping the taper angle constant for a particular width. In other words, the average thickness ($(t_o +t_i)/2$) is increased keeping the thickness of each layer proportional (1:1). Note that, for any given average thickness in the case of different widths, the thickness of either of the longitudinal edge is fixed and constant for different widths. With the increase in average thickness (measure of thickness in $xz$-plane), the number of turns decrease and pitch increases linearly as shown in Figs.~\ref{9a} and ~\ref{9b}, for the small
thickness edge and the large thickness edge, respectively.\\ Radius increases linearly and equally for both small thickness and large thickness edge, with increasing average thickness, due to increasing bending stiffness and as indicated by the parallel lines of both edges (see Fig.~\ref{9c}). The cone angle remains constant as the taper angle is maintained same (Fig.~\ref{9d}), independent of average thickness. Variation of the cone angle with aspect ratio, is in accordance with previous observations, which is, the cone angle decreases with increasing aspect ratio as taper angle is also changing with the aspect ratio (Fig.~\ref{7b}).

\noindent
Thus, it can be inferred that, for the thickness variation along the width, the deformation in $y$-direction is due to bending moment arising from the stiffness variation along that direction. Thus, the pitch so obtained increases continuously as the bending stiffness variation is also continuous. Radius for a particular average thickness remains the same as the same bending stiffness at that longitudinal section and the same strain is acting in that plane (initially $xz$-plane).
 The structure so obtained is a variation of surface of a helicoid with varying pitch and straight line generatrix with a cylindrical directrix whose parametric equation is given by \cite{kriv}.
\subsection{Thickness tapering along the length}
\begin{figure}[!htbp]
\centering
        \begin{subfigure}[b]{0.49\textwidth}
                \includegraphics[width=\linewidth]{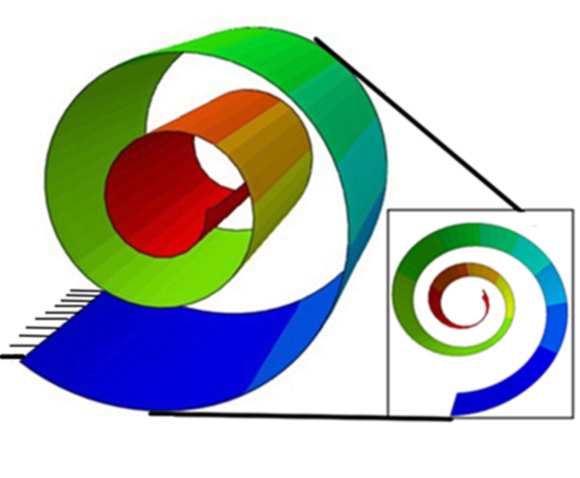}
				\put(-195,12){Fixed end}
                \put(-60,7){Front view}
                \caption{}
                \label{10a}
        \end{subfigure}
        \hfill
	\begin{subfigure}[b]{0.49\textwidth}
	\includegraphics[width=\linewidth]{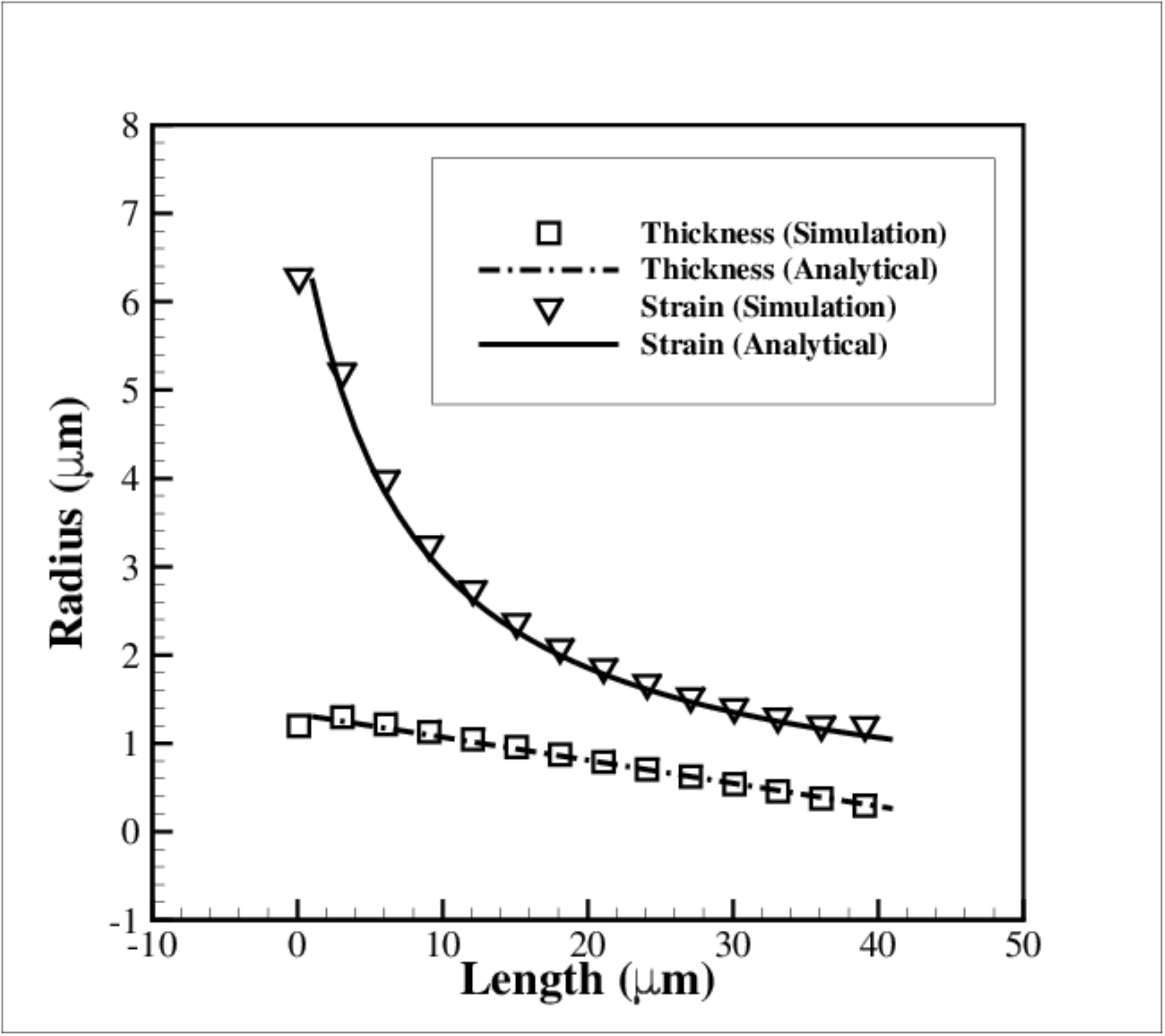}
	\caption{}
	\label{10c}
	\end{subfigure}
	\caption{a) Effect of tapered thickness and strain gradient along the length in a bilayer system and b) Evolution of radius with thickness variation ($50$ to $10$ nm) and strain variation ($0.75\%$ to $3\% $) (left to right, linearly)}
	\label{10}
\end{figure}

The effects of variation of thickness along the length are studied in this section. The initial configuration of the bilayer is depicted in Fig.~\ref{3b}. The shape obtained after straining to the bilayer, is that of a tube with increasing radius (free end to fixed end) (See Fig.~\ref{10a}). The radius thus obtained, is a function of thickness.\\
The FE model predicts variable radius, and is in very good agreement with the analytical solution (equation (1)) for a given thickness at a given position along the length, as shown in Fig.~\ref{10c}. The radius of curvature for a given thickness of a uniform thickness bilayer, was calculated using equation (1), obtaining different radius for different thickness of the uniform bilayer and plotted in Fig.~\ref{10c} along withthe  radius of curvature for a linearly tapered thickness along the length of the bilayer system. The FE analysis plot depicts the radius at a particular point of the tapered bilayer along the length.
\subsection{Effects of strain gradient along the width}
In this section, we present the results concerning the second methodology, i.e., simulations by strain engineering to fabricate conical films. First, we discuss the system with strain gradient along the width of the film and then along the length of the film (discussed in the next section). The values of strains at small strain edge and large strain edge for both the cases are shown in Figs.~\ref{14a} and \ref{14b}. Note that in this section, the thickness of the bilayer film is taken as 30 nm (each layer) and is uniform in all simulations for the mentioned strain gradient.\\
\begin{figure}[!hb]
\centering
        \begin{subfigure}[t]{0.49\textwidth}
                \includegraphics[width=\linewidth]{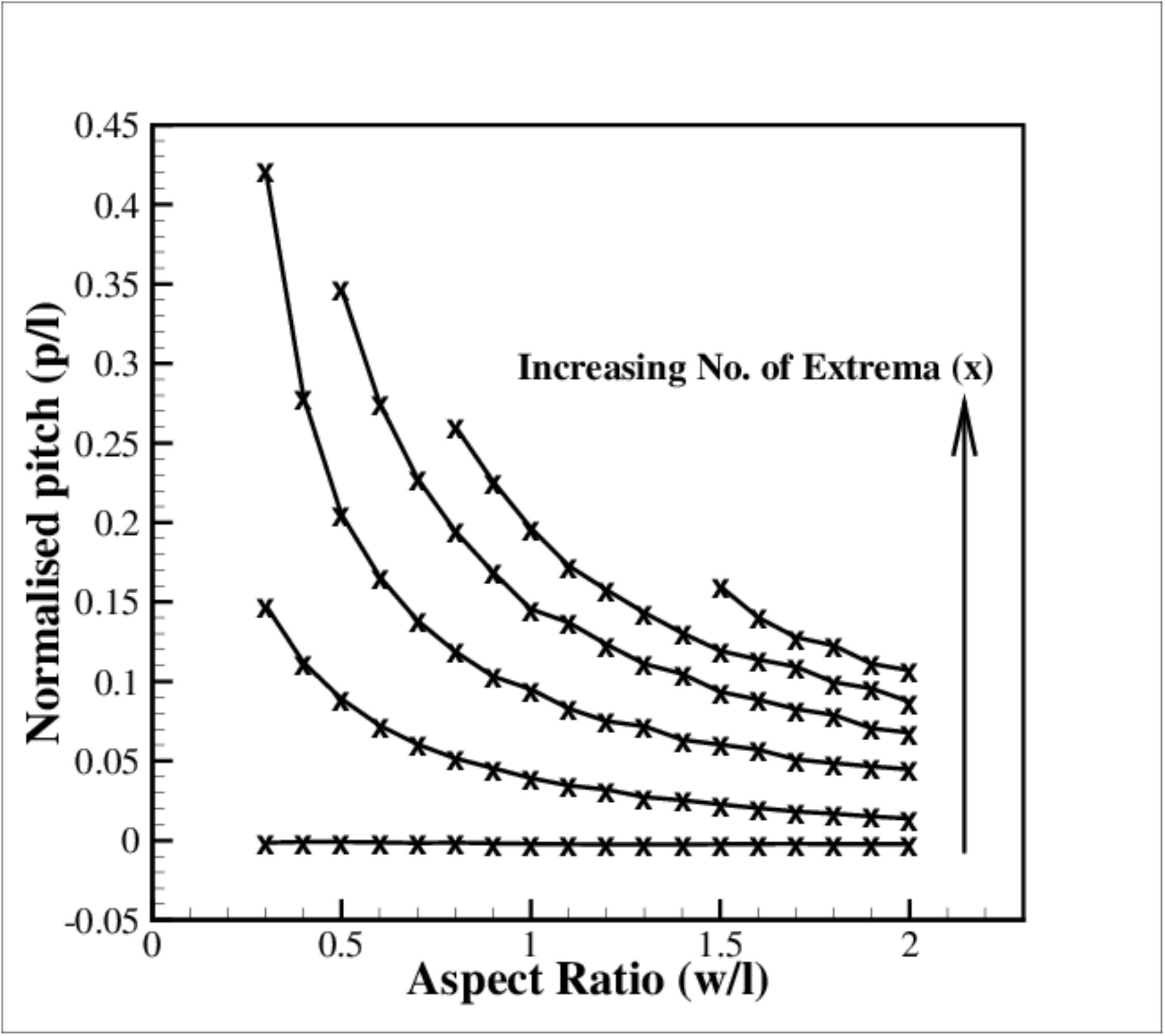}
                \caption{}
                \label{11a}
        \end{subfigure}%
        \hfill
        \begin{subfigure}[t]{0.49\textwidth}
                \includegraphics[width=\linewidth]{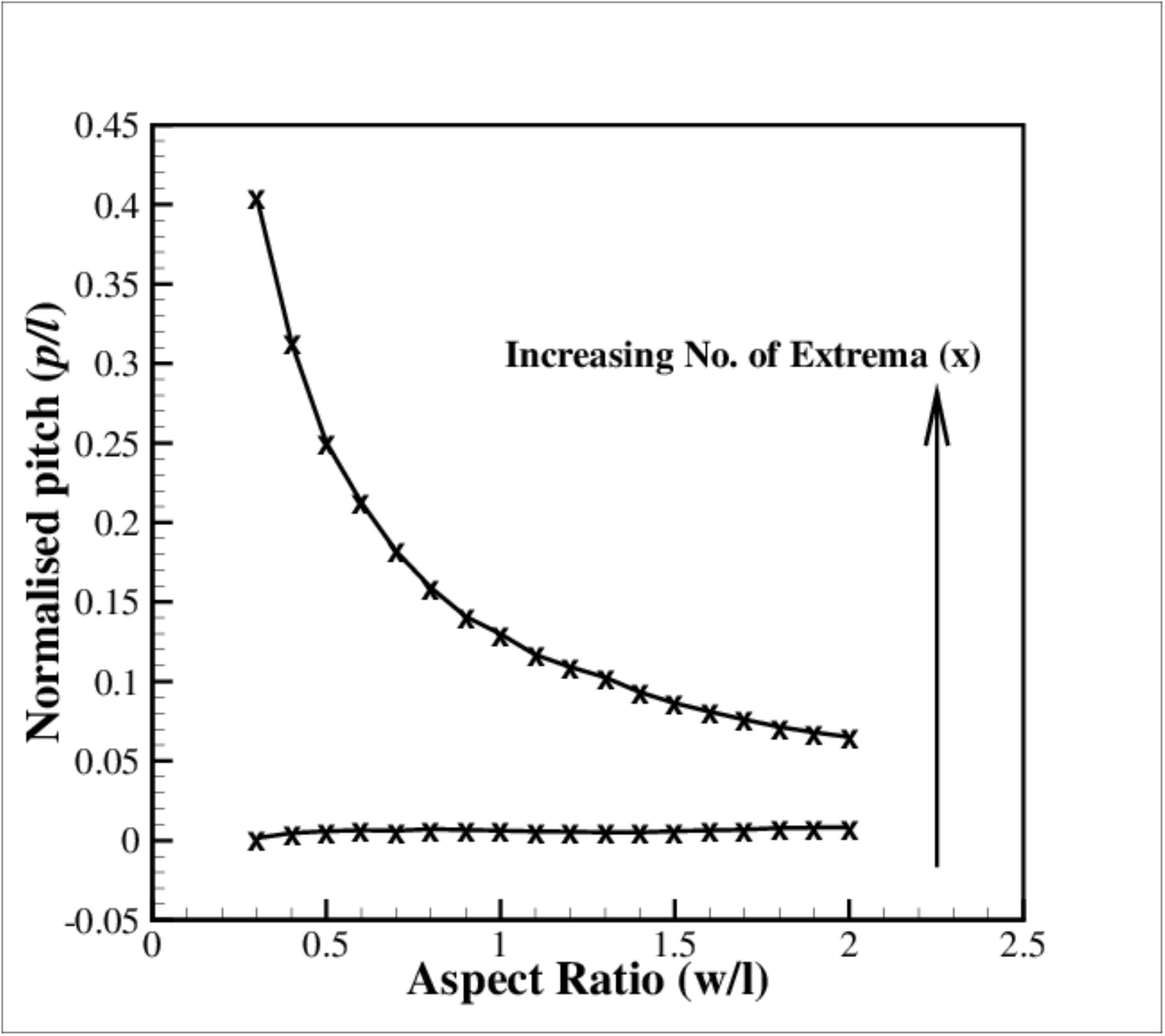}
                \caption{}
                \label{11b}
        \end{subfigure}%
        \hfill
        \begin{subfigure}[t]{0.49\textwidth}
                \includegraphics[width=\linewidth]{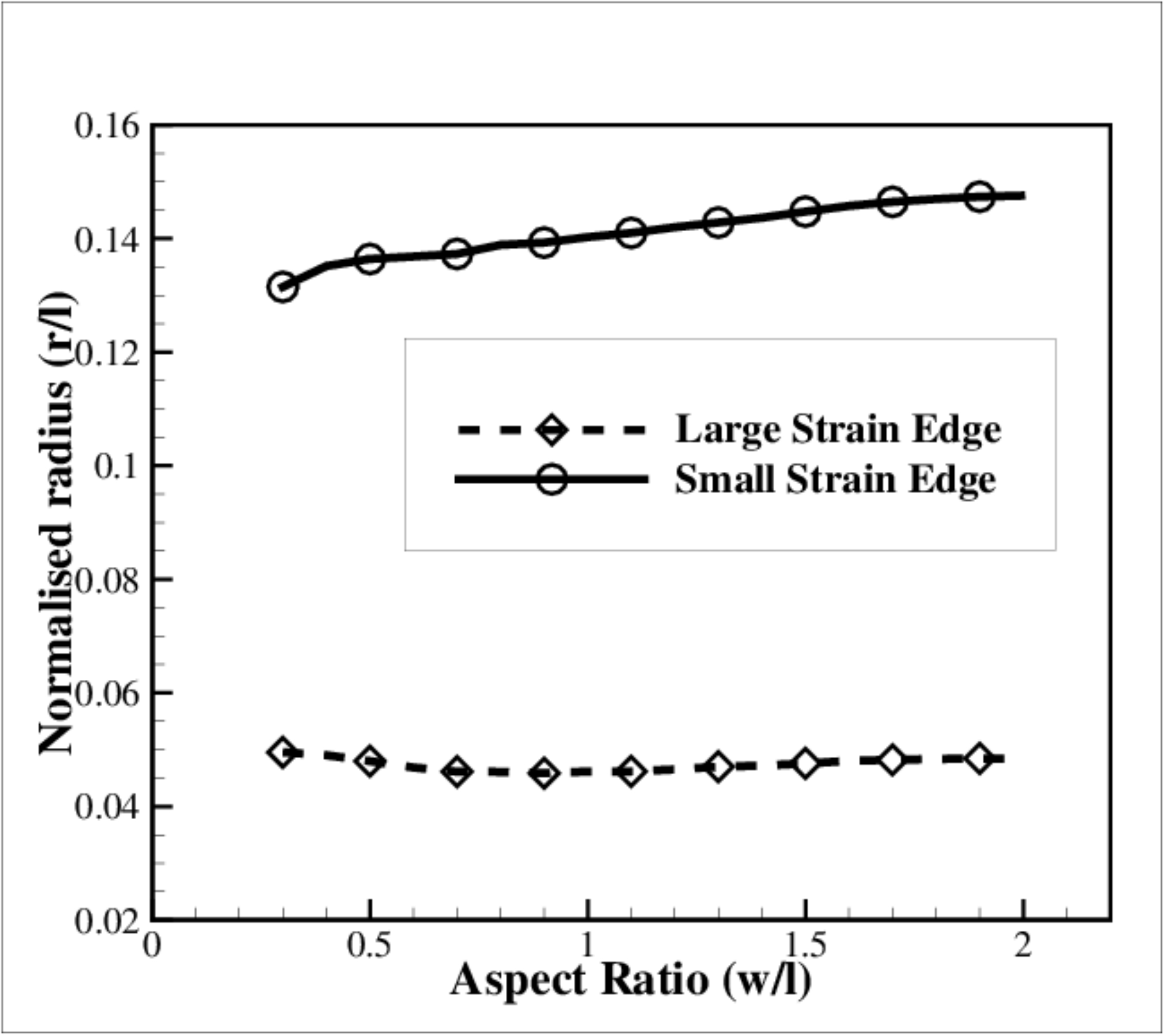}
                \caption{}
                \label{11c}
        \end{subfigure}%
        \hfill
        \begin{subfigure}[t]{0.49\textwidth}
                \includegraphics[width=\linewidth]{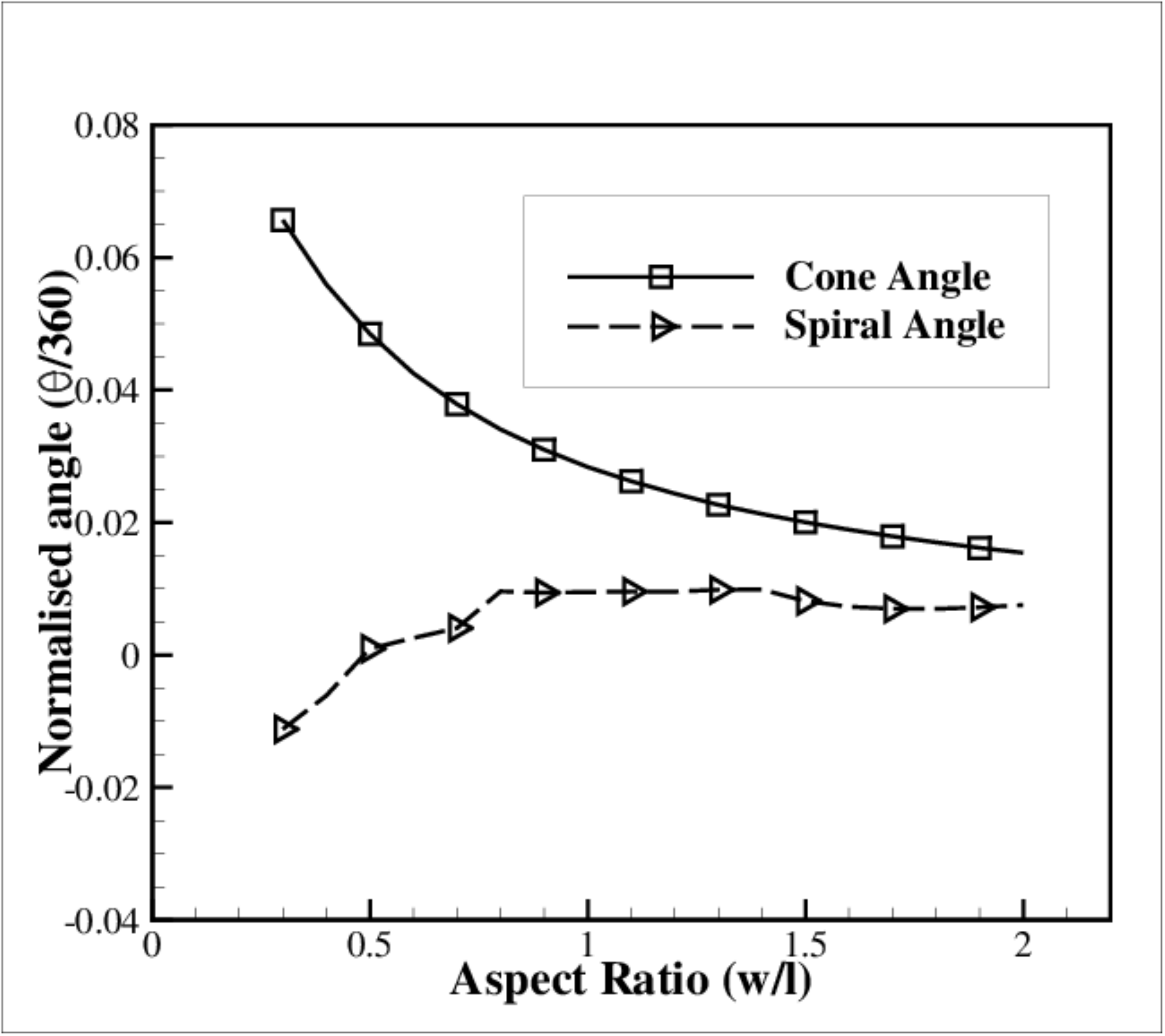}
                \caption{}
                \label{11d}
        \end{subfigure}%
        \caption{Strain gradient effect: a) Pitch variation for large strain edge, b) For small strain edge, c) Normalised radius ($r/l$) (referred by edge) and d) Normalised cone angle and normalise spiral angle}
        \label{11}
\end{figure}

\noindent Figs.~\ref{11a} and \ref{11b} show the variation of normalized pitch as a function of aspect ratio ($w/l$) of the film for large strain edge and small strain edge, respectively (0.5$\%$ strain at an edge is termed small strain edge and 3$\%$ at the other edge is termed as large strain edge). It can be observed that the number of extrema increases and the pitch decreases with increase in the aspect ratio ($w/l$). However, the edges are switched, i.e., the edge with higher strain curls in smaller radius. The effects of strain gradient along the width for different aspect ratios ($w/l$) as shown in Fig.~\ref{11} are similar to the effects due to thickness variation along the width as shown in Figs.~\ref{6} and \ref{7}.\\

\begin{figure}[!htbp]
\centering
        \begin{subfigure}{0.49\textwidth}
                \includegraphics[width=\linewidth]{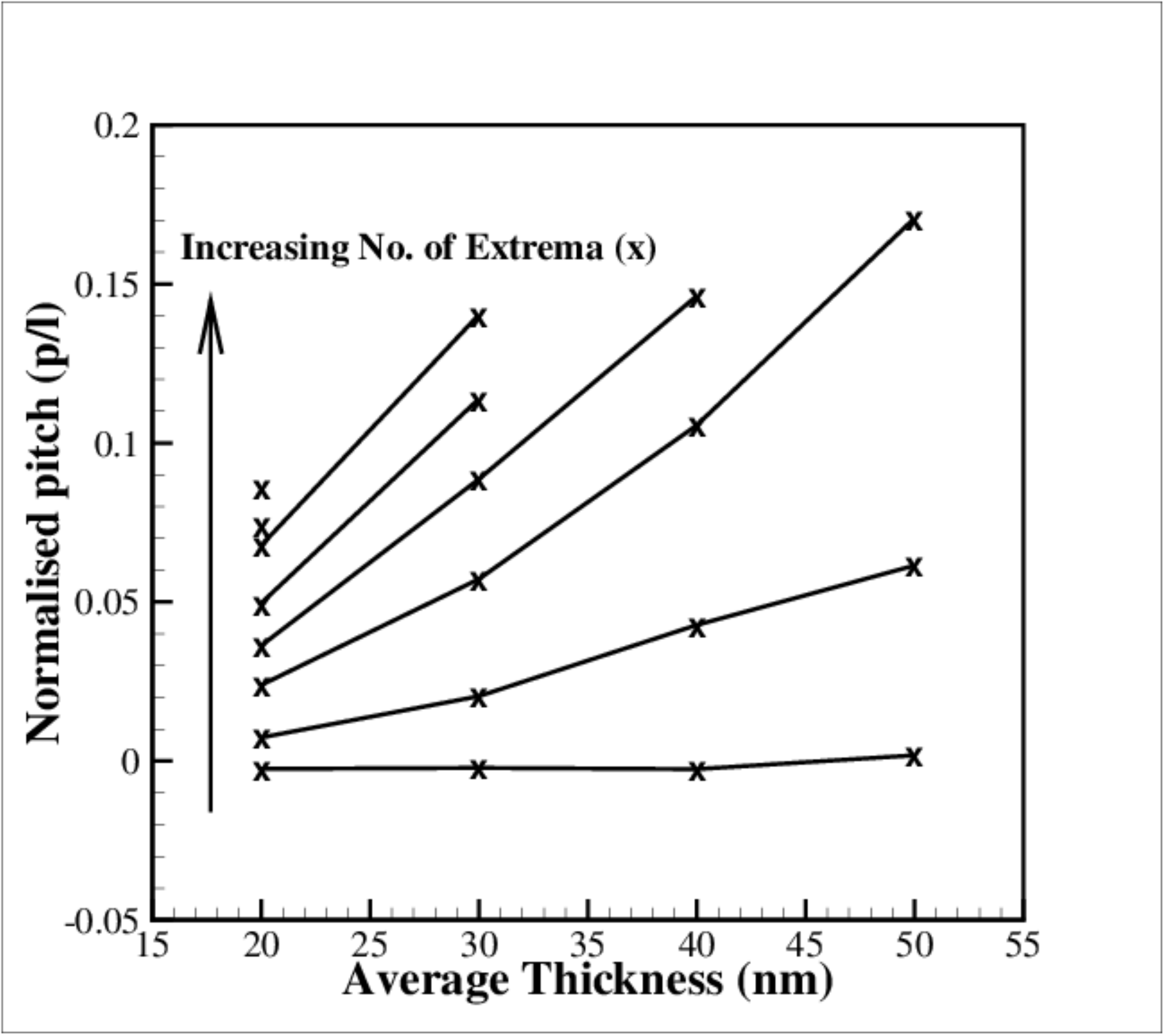}
                \caption{}
                \label{12a}
        \end{subfigure}%
        \hfill
        \begin{subfigure}{0.49\textwidth}
                \includegraphics[width=\linewidth]{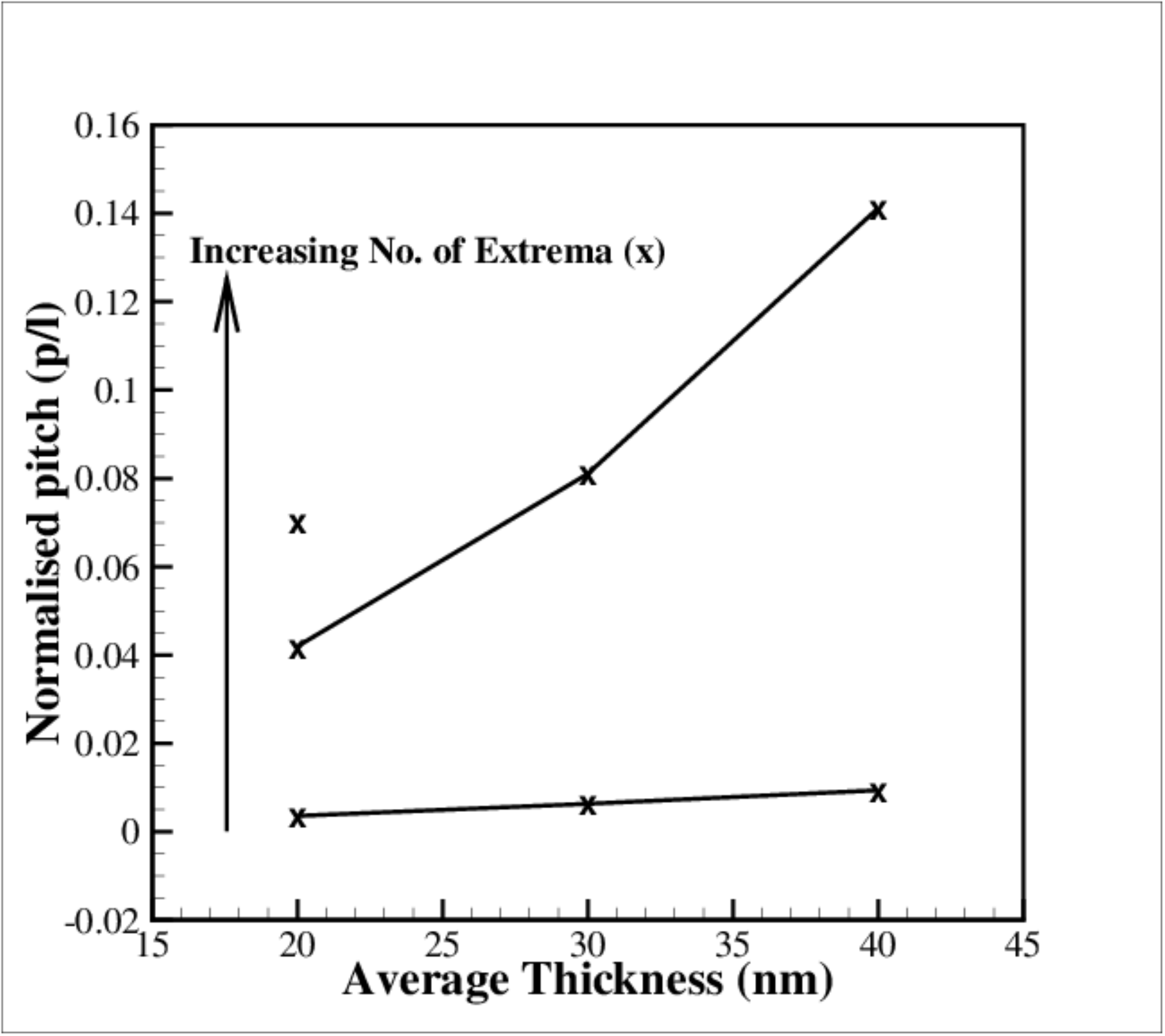}
                \caption{}
                \label{12b}
        \end{subfigure}%
        \hfill
        \begin{subfigure}{0.49\textwidth}
                \includegraphics[width=\linewidth]{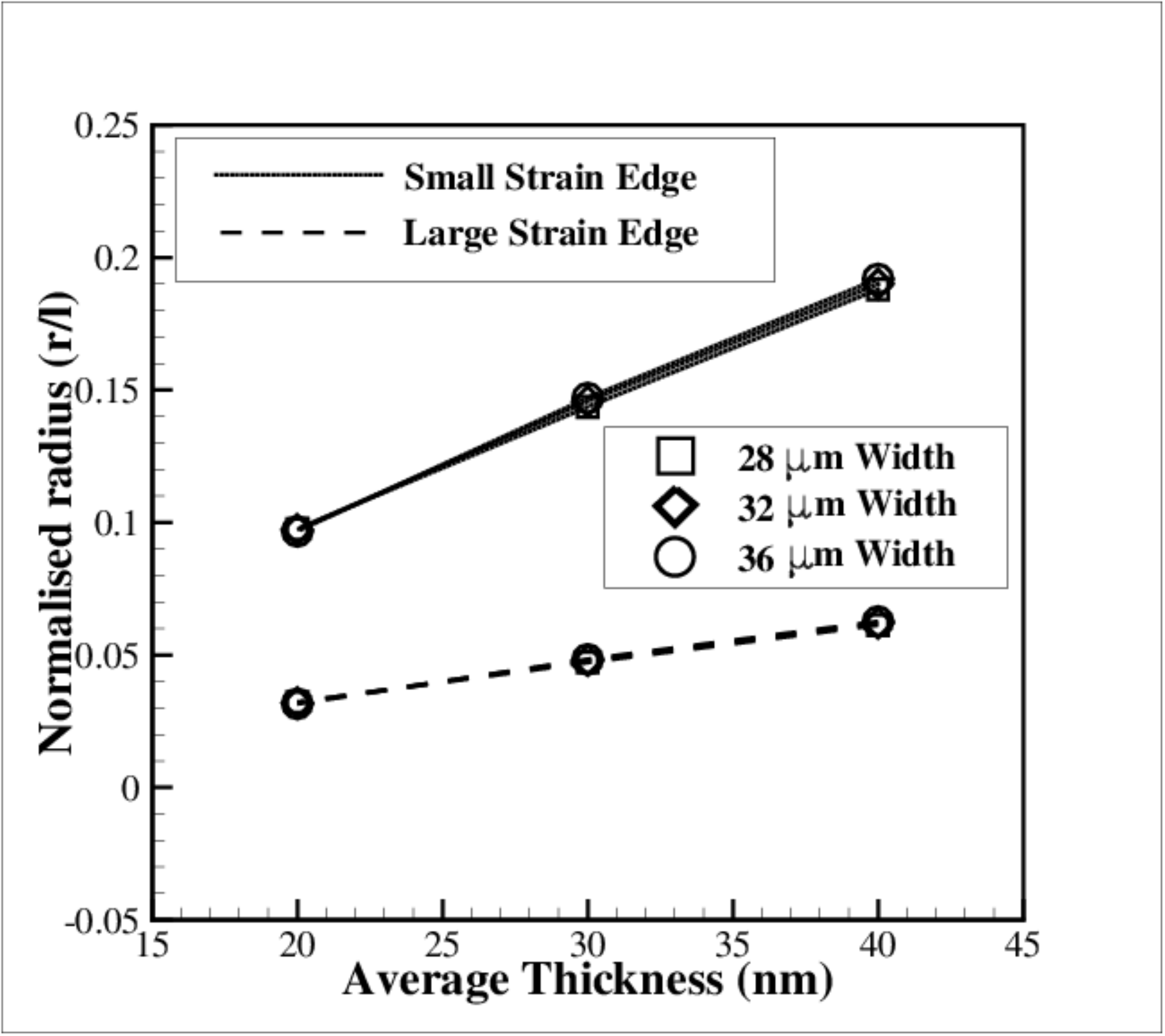}
                \caption{}
                \label{12c}
        \end{subfigure}%
        \hfill
         \begin{subfigure}{0.49\textwidth}
                \includegraphics[width=\linewidth]{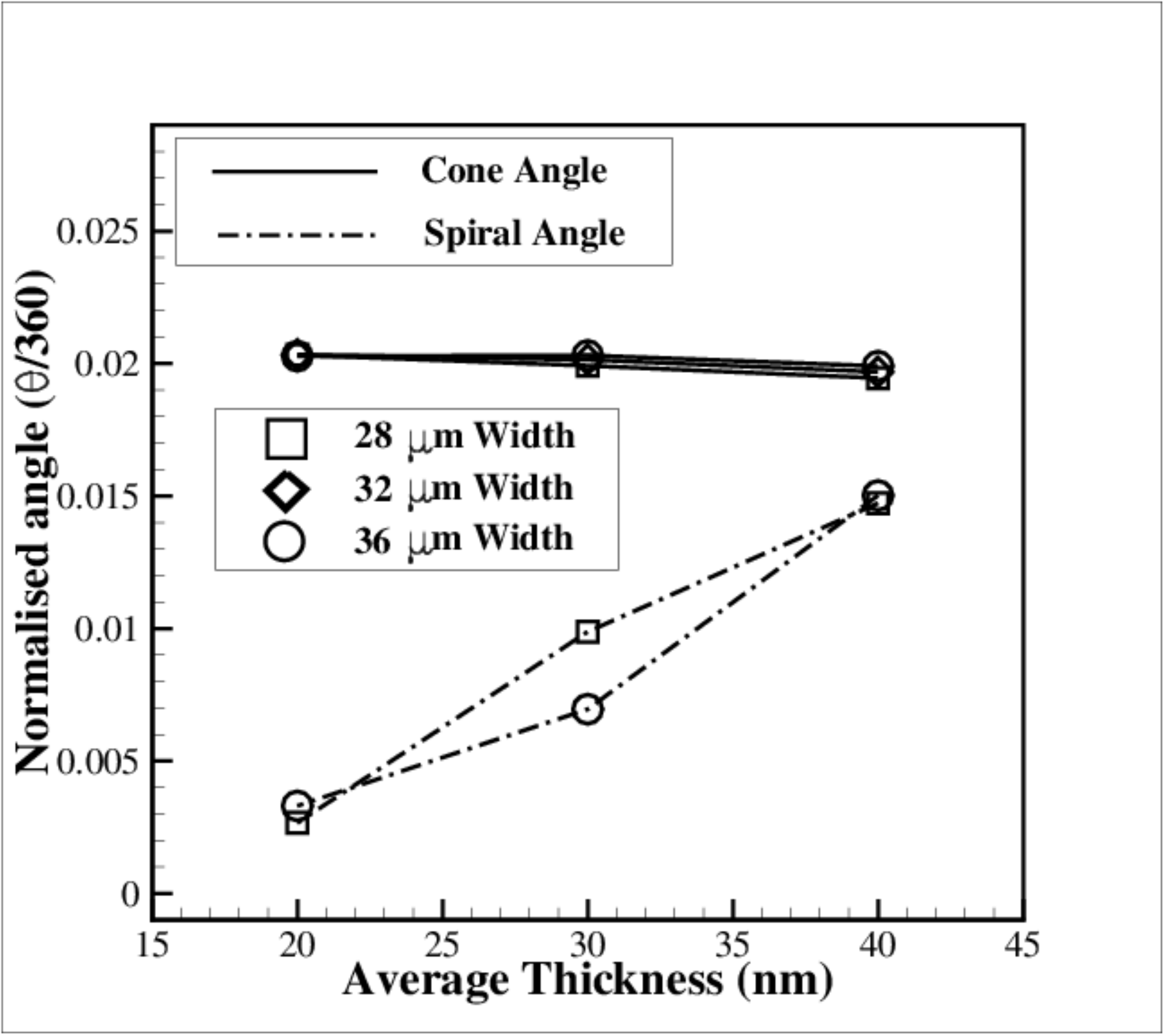}
                \caption{}
                \label{12d}
        \end{subfigure}%
        \caption{Average thickness effect (on strain variation model): a) Pitch variation for large strain edge, b) For small strain edge, c) Normalised radius ($r/l$) (referred by edge) and d) Normalised cone angle and  normalised spiral angle}\label{12}
\end{figure}
\begin{figure}[!htbp]
\centering
        \begin{subfigure}{0.49\textwidth}
                \includegraphics[width=\linewidth]{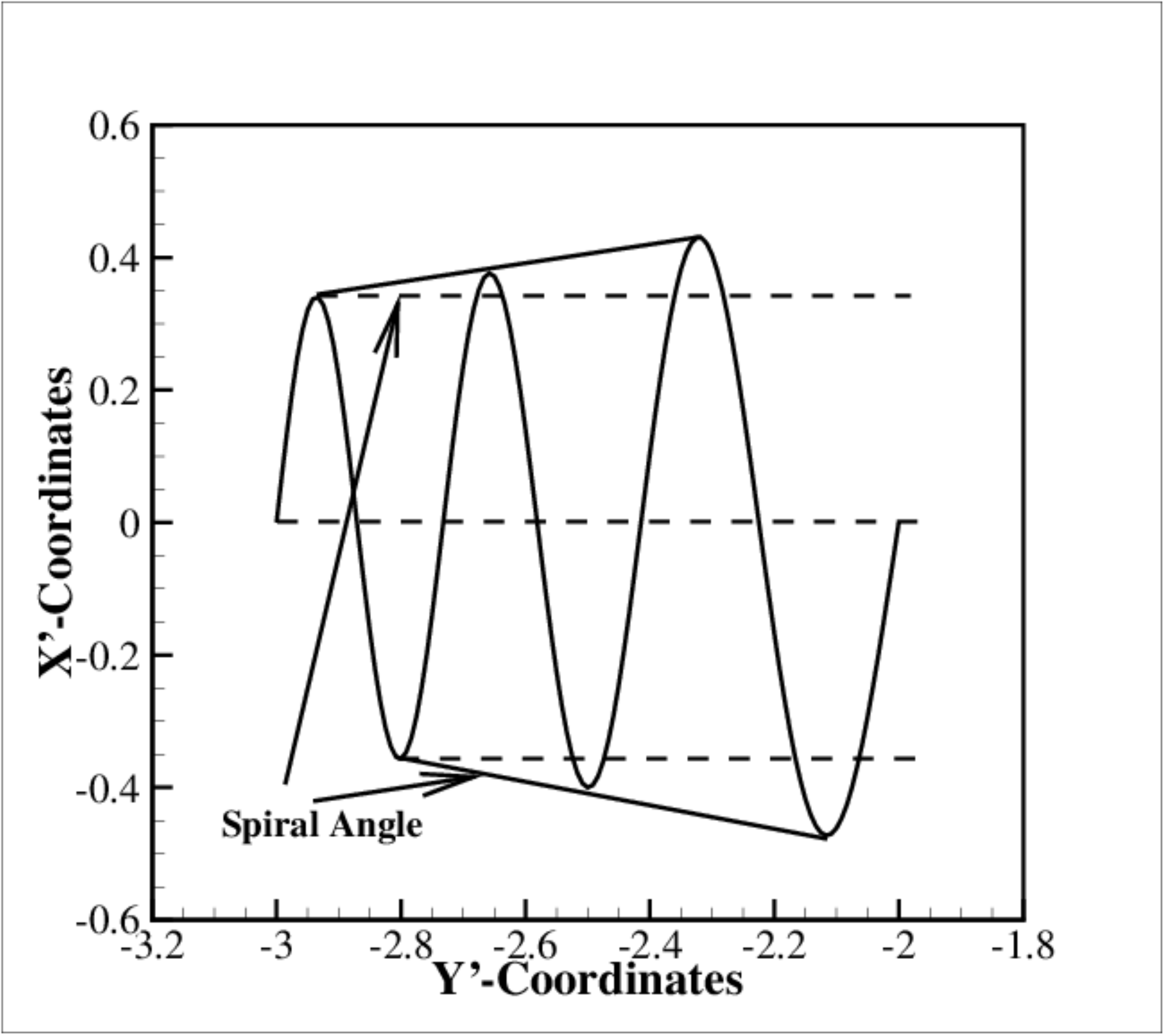}
                \caption{}
                \label{13a}
        \end{subfigure}%
        \hfill
        \begin{subfigure}{0.49\textwidth}
                \includegraphics[width=\linewidth]{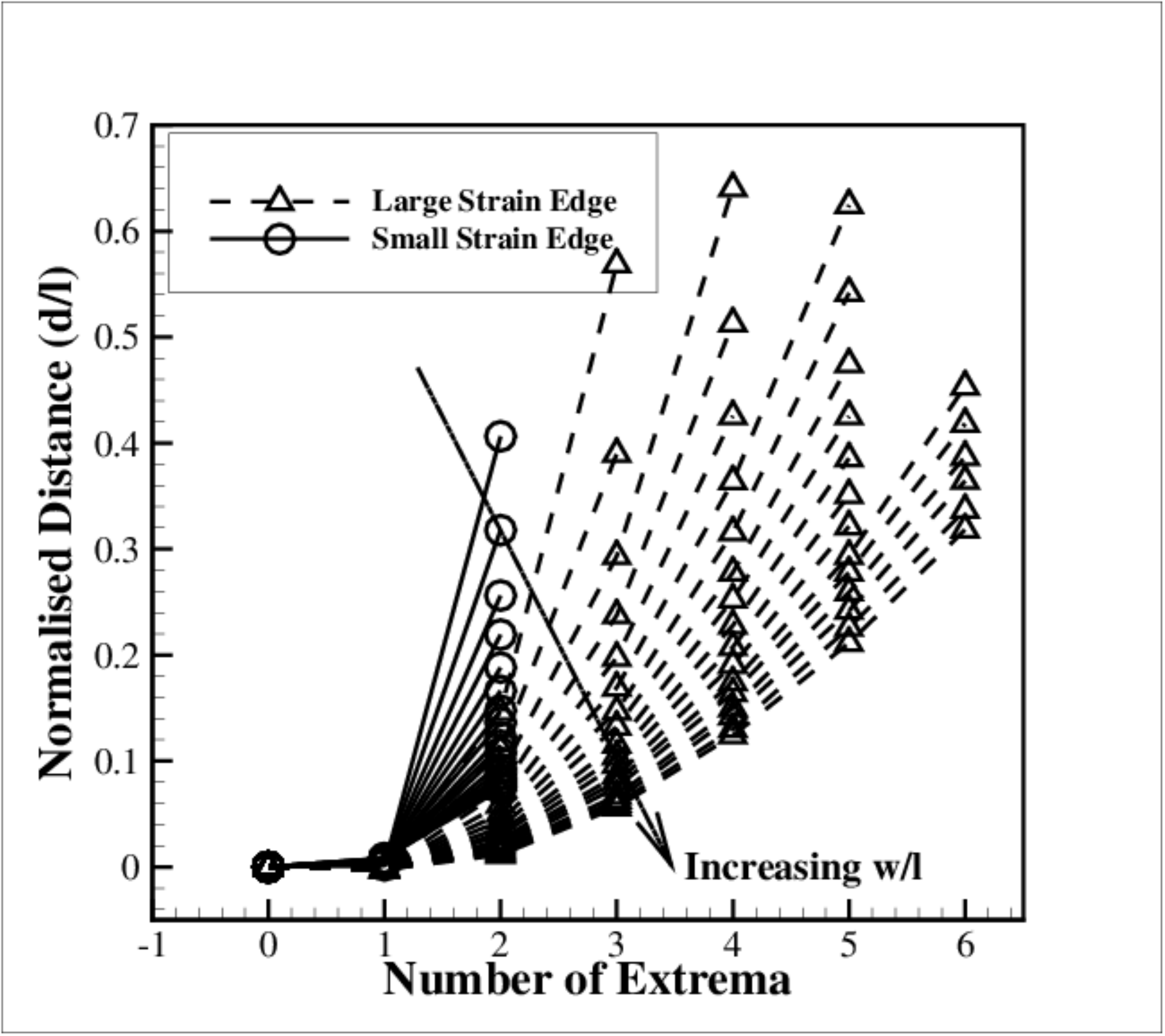}
                \caption{}
                \label{13b}
        \end{subfigure}%
\caption{a) Depiction of spiral angle and, b) Location of an extrema along median wave-line path $d$ for different aspect ratios ($w/l$) from fixed end to free end of an edge's projection plot}
\label{13}
\end{figure}

\noindent
Effect of average thickness on the normalized pitch for width 32 $\mu$m, is shown in Figs.~\ref{12a} and \ref{12b}. Here, the average thickness is the average of the thickness of top and bottom layers of the bilayer. Note that the number of turns and the pitch variation in this case are less compared to the case of film with uniform strain and thickness gradient along the width (see Figs.~\ref{9a} and \ref{9b}). Fig.~\ref{12c} shows the radius variation as a function of average thickness and Fig.~\ref{12d} shows the cone angle and spiral angle (spiral angle is the term used for angle formed by a linear increase in radius of a longitudinal section of the bilayer film from fixed end to free end) as a function of average thickness. The radius of the roll increases (see Fig.~\ref{12a}) with increase in average thickness due to increased bending stiffness for both the small strain and large strain edges. However, it is worth noting that the rate of change of radius with thickness is not the same for both the edges unlike the case of the thickness variation along the width (Fig.~\ref{9c}). This different rates of radius variation with the average thickness for both the edges (Fig.~\ref{12a}) may be attributed to the existence of spiral angle. \\
\noindent
Thus, comparative observations can be made with the thickness variation along the width, with an additional angle of spiral as depicted in Fig.~\ref{13a}. The spiral angle remains constant for three or more maxima as observed in Fig.~\ref{11d}, and thus, is independent of the strain gradient. Also, as seen in Fig.~\ref{12d}, the cone angle remains same since there is no change in the strain gradient, but the spiral angle increases with the increase in overall stiffness which in turn is due to the increase in the average thickness. Due to the spiral angle effect, only the initial radius (starting radius of helical spiral) (first turn on the conical tube), of various edges, is relevant for defining the rolled-up assembly.\\
Note that below three maxima of large strain edge, it is not possible to get a proper spiral angle, due to the effect of fixed boundary condition on the first maxima and minima. Also, beyond a total thickness of 80 nm of the bilayer, the edge with small strain does not roll in a complete turn, thus, some measurements are not possible for those cases (cone angle, radius of the small strain edge).\\
The spiral angle comes in effect due to the strain being bi-directional, hence, it can relax in the plane of the bilayer. It was not so in the case of thickness variation along the width as bending stiffness was limited to one direction, and is a fixed entity for the entire bilayer.\\
In conclusion, It was observed that the thickness variation along the width has a larger effect (with direct proportionality) on the radius, cone angle and pitch variation, as compared to the strain variation (with inverse proportionality).
\subsection{Effects of strain gradient along the length}
In this section, we present the results corresponding to the case of a bilayer film subjected to variation of strain along the length of the film. The values of strain at the fixed
and the free lateral edges are shown in Fig.~\ref{14b}. Note that the strain at the free edge is 3$\%$ and at the fixed edge is 0.75$\%$ with a linear variation. The simulation predicts
variable radius as shown in Fig.~\ref{10} showing a very good agreement with the analytical solution (equation (1)) (same procedure was followed in calculations as for thickness
taper along the length).\\
Here, it can be concluded that the thickness variation along the length has lower effects on the bilayer radius (with direct proportionality) compared with the strain variation along the length (with inverse proportionality).
\section{Conclusions}
We propose a novel method to fabricate conical tubes based on the same principle of out-of-plane deformation of bilayer thin films. It has been shown that by varying the thickness or strain along the width of the bilayer film, it is possible to obtain conical tubes of different geometrical properties. Detailed analysis on the effects of thickness variation and strain variation along the width is conducted to quantify the geometry of the thus formed cones. It was observed that the cone angle
is directly influenced by the taper angle and the strain gradient. The radius of the cone at a given section is influenced by thickness (direct proportionality) and strain (inverse proportionality) at that section. The radius remains constant for a given longitudinal section of uniform thickness in the case of thickness gradient along the width. But in the case of strain gradient along the width, the radius of a given longitudinal section shows an increase in radius from fixed end to free end, which is counter intuitive. The radius change along the width of the film results in the cone angle and the radius change along the length (discussed in the previous sentence) results in spiral angle. The spiral angle appears in the rolled-up geometry of the conical bilayer for the strain variation case alone, due to the non-equi-biaxial nature of the eigenstrains. In the case of thickness variation along the width, we do not observe a spiral angle formation due to the equi-biaxial state of strain. The geometry of the cone is described by helices of constant radius and variable pitch for the thickness variation. In the case of a uniform thickness bilayer film with strain gradient along the width, the helices will be of increasing radius with variable pitch. It has been observed that the strain (or thickness) variation along the length of the film results in the formation of spiral geometry.\\
The work presented here can be helpful in predicting the dimensions of the resultant geometrical features of the rolled-up tubes and cones for a given initial geometry of the film and the eigenstrain. Thus, the methodologies int his work may help designers to come up with novel fabrication strategies of micro and nano-scale three dimensional objects from two-dimensional geometries through strain and geometry engineering.
\bibliographystyle{elsarticle-num-names}
\bibliography{sample}
\end{document}